\documentclass[floats,floatfix,amssymb,prd,twocolumn,superscriptaddress,nofootinbib,preprintnumbers]{revtex4-2}

\usepackage{subcaption}
\usepackage{ragged2e}
\DeclareCaptionJustification{justified}{\justifying}
\makeatletter
\newcommand{\subsetsim}{\mathrel{\mathpalette\subset@sim\relax}}
\newcommand{\subset@sim}[2]{%
  \vtop{\offinterlineskip\m@th
    \ialign{\hfil##\cr
      $#1\subset$\cr\noalign{\kern0.5pt}\scalebox{0.9}{$#1\sim$}\cr
    }%
  }%
}
\makeatother

\usepackage{amssymb,amsmath,verbatim,mathtools,needspace,enumitem,etoolbox,graphicx,microtype,afterpage,bm}

\usepackage[dvipsnames, usenames]{xcolor}

\definecolor{linkcolor}{rgb}{0.0,0.3,0.5}
\usepackage{booktabs}
\usepackage[unicode, colorlinks=true, linkcolor=linkcolor, citecolor=linkcolor, filecolor=linkcolor,urlcolor=linkcolor, pdfusetitle]{hyperref}
\usepackage[all]{hypcap}
\usepackage[T1]{fontenc}
\usepackage[utf8]{inputenc}
\usepackage{tabularx}
\usepackage{float}
\usepackage{multirow}
\usepackage{pifont}
\usepackage{lmodern}

\allowdisplaybreaks
\usepackage{tikz}
\usetikzlibrary{shapes.misc, positioning, arrows.meta}
\usepackage{framed}
\usepackage{hyperref}
\hypersetup{colorlinks, citecolor=bluscuro, linkcolor=black, urlcolor=bluscuro}
\definecolor{rossos}{cmyk}{0,1,1,0.55}
\definecolor{bluscuro}{rgb}{0.15, 0.2, .85}
\definecolor{bluchiaro}{cmyk}{1,.3,0.,0.1}
\definecolor{ForestGreen}{rgb}{0.13, 0.55, 0.13}
\definecolor{TLGreen}{RGB}{50, 164, 49}
\definecolor{TLOrange}{RGB}{231,180,22}
\definecolor{TLRed}{RGB}{204,50,50}

\newcommand{\TLBullet}[1]{\raisebox{-5pt}{\scalebox{0.23}{\begin{tikzpicture}\shadedraw[rounded corners=15pt, top color=gray!84!black,bottom color=black, line width=.6pt] (0,0) rectangle ++(6,2); \ifthenelse{#1=1}{\draw[fill=green,line width=1.pt]  (1,1) circle(.75cm);}{\draw[fill=green!35!black,line width=1.pt]  (1,1) circle(.75cm);}\ifthenelse{#1=2}{\draw[fill=yellow,line width=1.pt]  (3,1) circle(.75cm);}{\draw[fill=yellow!60!black,line width=1.pt]  (3,1) circle(.75cm);}\ifthenelse{#1=3}{\draw[fill=red,line width=1.pt]  (5,1) circle(.75cm);}{\draw[fill=red!50!black,line width=1.pt]  (5,1) circle(.75cm);}\end{tikzpicture}}}}

\def\d{{\mathrm{d}}}

\newcommand{\bs}{\begin{subequations}}
\newcommand{\es}{\end{subequations}}

\newcommand{\be}{\begin{equation}}
\newcommand{\ee}{\end{equation}}
\renewcommand{\d}{{\rm d}}

\def\lsim{\mathrel{\rlap{\lower4pt\hbox{\hskip0.5pt$\sim$}}
    \raise1pt\hbox{$<$}}}         
\def\gsim{\mathrel{\rlap{\lower4pt\hbox{\hskip0.5pt$\sim$}}
    \raise1pt\hbox{$>$}}}         

\usepackage{siunitx}
\DeclareSIUnit \parsec {pc}
\DeclareSIUnit \arcsecondfull {arcsec}
\DeclareSIUnit \year{yr}
\DeclareSIUnit \day{day}
\DeclareSIUnit \hour{hr}
\DeclareSIUnit \radiant{rad}
\DeclareSIUnit \degfull{deg}
\DeclareSIUnit \erg {erg}
\DeclareSIUnit \Lsun {L_\odot}
\DeclareSIUnit \Msun {M_\odot}
\DeclareSIUnit \AstroUnit {au}

\newcommand{\dd}[1]{{\rm d}#1}

\usepackage{nicefrac}

\newcommand{\unipd}{Dipartimento di Fisica e Astronomia ``G. Galilei'', Università degli Studi di Padova, via Marzolo 8, I-35131 Padova, Italy}
\newcommand{\infnpd}{INFN, Sezione di Padova, via Marzolo 8, I-35131 Padova, Italy}

\usepackage{physics}
\usepackage{acronym}

\newcommand{\uc}{\mathrm{c}}

\newcommand{\bae}[1]{\begin{align} #1 \end{align}}

\newcommand{\bme}[1]{\begin{multline} #1 \end{multline}}

\definecolor{MONZA}{HTML}{CF000F}
\definecolor{DARKBLUE}{HTML}{00008b}
\definecolor{DARKMAGENTA}{HTML}{8b008b}

\usepackage{soul}

\begin{document}

\title{
QCD Crossover Transfer Functions for 
\\
Scalar-Induced Gravitational Waves in the PTA Band
}

\author{Gabriele Franciolini}
\email{gabriele.franciolini@unipd.it}
\affiliation{\unipd}
\affiliation{\infnpd}

\author{Xavier Pritchard}
\email{xavpritchard@outlook.com}
\affiliation{Astronomy Centre, University of Sussex,
Falmer, Brighton, BN1 9QH, UK}

\author{Yuichiro Tada}
\email{ytada@u-fukui.ac.jp}
\affiliation{Department of Applied Physics, University of Fukui,
Bunkyo 3-9-1, Fukui 910-8507, Japan}

\date{\today}

\begin{abstract}
Pulsar timing array (PTA) collaborations have reported evidence for a stochastic gravitational wave (GW) background in the nHz band. Should scalar-induced GWs, sourced at second order by enhanced primordial curvature perturbations, contribute to this signal, a coincidence of scales makes them directly sensitive to the softening of the equation of state around the QCD crossover. We solve the tensor and scalar equations of motion across the Standard Model (SM) thermal history and tabulate the transfer functions for direct use in present and future PTA analyses. We show that this SM effect modifies the height of the induced spectrum by up to $\approx55\%$ across the PTA band relative to the radiation-domination expectation, with either sign depending on whether the source modes cross the horizon before or after the crossover. Fitting the NANOGrav 15-year data with a broken-power-law curvature power spectrum, we find that including the crossover shifts the inferred peak amplitude and scale by an amount that could already be relevant for the comparison with primordial black hole overproduction bounds. The importance of this SM effect will grow as the statistical uncertainty on the amplitude and scale shrinks with future, more sensitive PTA datasets, at which point neglecting it could significantly bias the inference.
\end{abstract}

\maketitle

{
  \hypersetup{linkcolor=black}
}
\hypersetup{linkcolor=bluscuro}


\acrodef{GW}{gravitational wave}
\acrodef{SIGW}{scalar-induced gravitational wave}


\section{Introduction}\label{intro}

PTA collaborations have reported evidence for a
common-spectrum, spatially correlated signal consistent with a stochastic
GW background in the nHz band~\cite{NANOGrav,NANOGrav1,EPTA,EPTA1,EPTA2,PPTA,PPTA1,PPTA2,CPTA,Miles:2024seg}. A cosmological,
rather than astrophysical, origin remains on the table~\cite{Madge:2023cak,NANOGrav:2023hvm,EPTA:2023xxk,Figueroa:2023zhu,Ellis:2023oxs,Caprini:2024lxj,Moore:2021ibq}, and
scalar-induced GWs (SIGWs)~\cite{Tomita:1967wkp,Matarrese:1993zf,Acquaviva:2002ud,Mollerach:2003nq,Ananda:2006af,Baumann:2007zm,Domenech:2021ztg}, sourced at second order in
perturbation theory whenever the curvature power spectrum is enhanced on
the scales probed by the array, are among the most studied candidates. The same enhanced curvature perturbations are often invoked in models producing a substantial abundance of primordial black holes (PBHs)~\cite{Zeldovich:1967lct,Hawking:1971ei,Carr:1974nx,Carr:1975qj,Chapline:1975ojl,Byrnes:2025tji}, and their overproduction leads to tight bounds on the maximal SIGW amplitude.
Reconstructing the underlying curvature spectrum from the data, however,
requires a transfer function which accounts for the SM thermal history. 

It is instructive to consider the frequency ($f\equiv k/2 \pi$)
of modes that entered the Hubble scale at temperature $T$. This is
\begin{align}\label{eq.fre}
    f_{\cal H}& =  \frac{{\cal H}}{2\pi}
    \approx
    \SI{26}{nHz}
    \left(\frac{g_{*\rho}}{100} \right)^{\frac12}
    \!\left( \frac{g_{*s}}{100} \right)^{\!-\frac13}
    \left (\frac{T}{\rm GeV} \right ).
\end{align}
In the previous relation, we introduced the conformal Hubble scale
${\cal H}$, and $(g_{*\rho},g_{*s})$ which are the effective numbers of degrees of freedom for the relativistic energy and entropy densities.
From this relation, we immediately see that
nHz frequency modes re-entered the Hubble sphere when the plasma temperature was near the
QCD crossover temperature  $T_{\rm QCD} \simeq 
0.15\,\text{GeV}$.
This means that a SIGW
interpretation of the PTA signal is inevitably probing the epoch of radiation-domination (RD) with the largest known SM departure from $w=1/3$.

In this paper, we solve the coupled tensor--scalar system on the exact
SM background across the QCD crossover, assemble the result
into tabulated, ready-to-use transfer functions, and quantify their impact
on the induced spectrum and on a fit to the NANOGrav 15-year data. 
The
paper is organized as follows. Sec.~\ref{sec:eos} sets up the thermal
history and the softened equation of state. Sec.~\ref{sec:sigw} derives
the induced tensor spectrum and its numerical evaluation, from the
sub-horizon WKB continuation to the tabulated kernels used in the rest of
the paper. Sec.~\ref{sec:results} presents the resulting GW spectra assuming representative 
scale-invariant and peaked curvature power spectra. Sec.~\ref{sec:pta}
applies these transfer functions to the NANOGrav 15-year data. We
conclude in Sec.~\ref{sec:conclusions}, and comment on the extension to
non-Gaussian curvature perturbations in the Appendix.

\section{The QCD crossover era}\label{sec:eos}

The thermodynamics of the early universe plasma may be described in terms of the total energy density, $\rho(T)$, and entropy density, $s(T)$, given by~\cite{Husdal:2016haj,Saikawa:2018rcs}.
\bae{
    \rho(T)=\frac{\pi^2T^4}{30}g_{*\rho}(T) \qc s(T)=\frac{2\pi^2T^3}{45}g_{*s}(T).
} 
Using these quantities, we may derive the main thermodynamic inputs of our calculation, namely, the equation-of-state parameter, $w(T)$, and the sound speed squared, $c_s^2(T)$, defined as
\bae{
    w(T)=\frac{p(T)}{\rho(T)} \qc c_s^2(T)=\pdv{p}{\rho}=\frac{p'(T)}{\rho'(T)},
}
where in order to describe the plasma pressure, $p(T)$, we make use of the relation $p(T)=Ts(T)-\rho(T)$.

Phase transitions in the early universe induce non-trivial, temperature-dependent variations
of $\rho(T)$ and $s(T)$, and therefore $w(T)$. Within the SM, the first departure from pure radiation occurs near the electroweak crossover at $T\simeq160\,\text{GeV}$, where electroweak symmetry breaking and the Higgs mechanism take place \cite{Englert:1964et,Higgs:1964pj,Weinberg:1967tq}. Thermodynamic quantities, including $w(T)$, have been computed for this epoch \cite{Laine:2015kra}. At a lower temperature, the plasma undergoes the QCD chiral crossover at $T\simeq0.15\,\text{GeV}$ \cite{Aoki:2006we,HotQCD:2018pds,Borsanyi:2020fev}, which leads to the largest known change in the radiation plasma equation of state. To describe the thermodynamics across this era, we use a cubic spline interpolation of modern lattice QCD results~\cite{borsanyi1} (see also Refs.~\cite{Borsanyi:2013bia,HotQCD:2014kol,Abuali:2025tbd,Borsanyi:2025dyp}). We note that the equation of state during the QCD era is sensitive to several effects beyond the SM, for example dark QCD \cite{Feng:2026xwp} or large lepton-flavor asymmetries~\cite{Gonin:2026tyu}. Finally, near $T\simeq10^{-3}\,\text{GeV}$, neutrino decouple and $e^-e^+$ annihilation occurs, producing a smaller, but still cosmologically relevant, change in $w(T)$.

After neutrino decoupling and $e^-e^+$ annihilation at $T \lesssim 10^{-3}\,\text{GeV}$, 
the plasma is no
longer a single thermal bath: neutrinos free stream at their own temperature
$T_\nu \ne T$, and pressure and energy density must be combined sector by
sector, weighted by energy density. This guarantees $w, c_s^2 \le 1/3$
everywhere, as required for a plasma composed of sectors that are each at
most radiation-like.
Although we do not consider lower frequency modes (i.e., $f\lesssim10^{-1}\,\text{nHz}$) as they are not observable by PTAs, we note that these free-streaming neutrinos have a damping effect on tensor modes \cite{Weinberg:2003ur,Watanabe:2006qe,Saikawa:2018rcs}, which we neglect here.

The background expansion is obtained by integrating the Friedmann and
continuity equations,
\begin{equation}
\mathcal H' = -\frac{1+3w}{2}\,\mathcal H^2 ,
\quad
3 M_{\rm Pl}^2 \mathcal H^2 = a^2 \rho ,
\label{eq:bg}
\end{equation}
as a coupled system for $T(\eta)$ and the scale factor $a(\eta)$, anchored
deep in the radiation era and normalized to $a=1$ today at
$T_0 = 2.725\,$K. Here $M_{\rm Pl}$ denotes the reduced Planck mass and a
prime denotes the derivative with respect to conformal time $\eta$.

\begin{figure}[t!]
\centering
\includegraphics[width=0.48\textwidth]{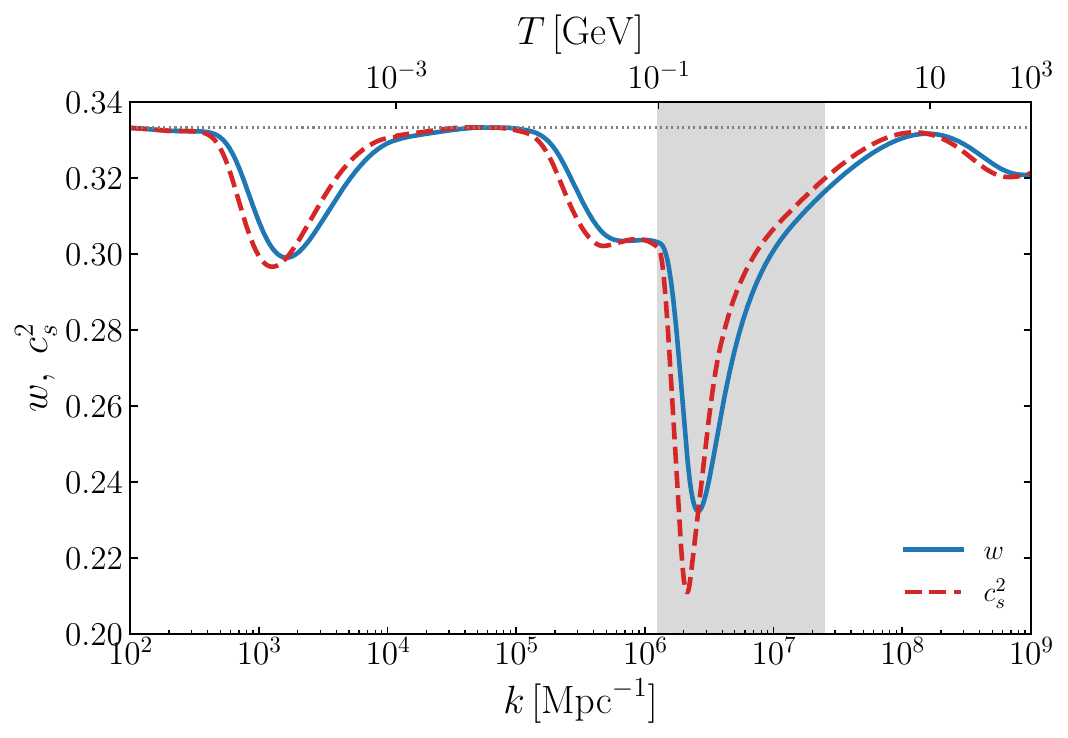}
\caption{Equation of state $w$ (solid) and sound speed squared $c_s^2$
(dashed) as functions of the comoving wavenumber $k$ crossing the horizon,
$k=\mathcal H$, with the corresponding plasma temperature on the top axis.
The dotted line marks the radiation value $1/3$. The features correspond,
from left to right, to electron--positron annihilation, the QCD crossover,
and the approach to the electroweak scale. The gray region indicates the current PTA sensitivity band
$f \in [1/T_{\rm obs}, 20/T_{\rm obs}]$, with $T_{\rm obs}$ being the current observation time of the datasets.}
\label{fig:eos}
\end{figure}

Figure~\ref{fig:eos} shows the resulting $w$ and $c_s^2$ as functions of the
comoving wavenumber $k$ of the mode crossing the horizon,
$k = \mathcal H(\eta)$, with the corresponding plasma temperature on the top
axis. Across the SM thermal history, there are several dips below the radiation value $1/3$: at the QCD crossover the minimum is $w \approx 0.23$, and at electron--positron annihilation $w \approx 0.3$, with a milder feature appearing toward the electroweak scale.
The QCD dip is the deepest SM feature and coincides with the
PTA sensitivity band, shown in gray.

\section{Second-order induced GWs}\label{sec:sigw}

\subsection{Equations of motion and induced spectrum}

Working in momentum space, the second-order tensor perturbation obeys~\cite{Tomita:1975kj,Matarrese:1993zf,Acquaviva:2002ud,Mollerach:2003nq,Ananda:2006af,Baumann:2007zm}
\begin{align}
    \Lambda_\eta\left [
    a(\eta)h _{\vec k}(\eta)
    \right]
    = 4 a(\eta) \int \frac{\d^3 q}{(2 \pi)^{3}}
    q_i q_j
    e_{ij}({\vec k })
    {\cal S}_{\vec k}(\eta),
\label{eq: EoM_gw}
\end{align}
where $h_{\vec k}$ is the tensor perturbation, $e_{ij}({\vec k })$ is the GW polarization tensor and we define the differential
operator
\begin{equation}\label{eq:Lammbda_eta}
     \Lambda_\eta  \equiv
     \frac{\d^2}{\d \eta^2} + k^2 - \left ( \frac{1-3w(\eta)}{2}  \right)
    {\cal H}^2 ,
\end{equation}
while the source term takes the form
\begin{align}
&{\cal S}_{\vec k} = 2\Phi_{\vec q}  \Phi_{\vec k -\vec q}
+ \frac{4}{3(1+w)}
\!\left(\! \frac{\Phi'_{\vec q}}{\mathcal{H}} + \Phi_{\vec q}\!\right)\!
\!\left(\! \frac{\Phi'_{\vec k -\vec q}}{\mathcal{H}}  + \Phi_{\vec k -\vec q}  \!\right)\!,
\end{align}
in terms of the conformal Newtonian gauge potential $\Phi$. We assume linar tensor perturbations from inflation to be negligible. 

The evolution of the scalar perturbations must account for the softening of
the equation of state around the QCD era. We therefore solve numerically the
Bardeen equation (see, e.g., Ref.~\cite{Mukhanov:2005sc})
\bme{\label{eq:phievo}
    \hat{\Phi}_{\vec k}''(\eta)+3{\cal H}(1+c_s^2)\hat{\Phi}_{\vec k}'(\eta) \\
	+ \left [c_s^2k^2+3{\cal H}^2(c_s^2-w) \right ]\hat{\Phi}_{\vec k}(\eta)=0.
}
Here, the quantum operator $\hat{\Phi}_{\vec k}(\eta)$ is decomposed into the transfer function
$\Phi_k(\eta)$ and the primordial perturbation $\hat{\psi}_{\vec k}$ as
$\hat{\Phi}_{\vec k}(\eta)=\Phi_k(\eta)\hat{\psi}_{\vec k}$. The latter is
controlled by the gauge-invariant super-Hubble curvature perturbation
$\hat{\zeta}_{\vec k}$ as $\hat{\psi}_{\vec k}=-2\hat{\zeta}_{\vec k}/3$.
Therefore, the initial condition adopted for the solution of the transfer
function is given by $\Phi_k(\eta)\to1$ and $\Phi_k'(\eta)\to0$ for
$\eta\to0$, which follows from the assumption of a perfect radiation fluid
dominating well before Hubble crossing, compatible with the SM thermal history.

The current abundance of the induced background is obtained by accounting for
the propagation as free GWs after emission, whose present energy density is
sensitive to deviations from an exactly radiation-dominated expansion through
the time dependence of $g_{*\rho}$ and $g_{*s}$. One
finds~\cite{Espinosa:2018eve,Kohri:2018awv,Domenech:2021ztg}
\begin{equation}
\label{eq:OmegaGW}
    \Omega_{\rm GW}(k)h^2
    = \Omega_{r,0} h^2\, c_g(k)\,
\overline{ {\cal P}_h(k,\eta_{\rm c})} ,
\end{equation}
with
\begin{equation}
    c_g(k) \equiv
    \left (\frac{a_{\rm c}{\cal H}_{\rm c}}{a_{\rm f}{\cal H}_{\rm f}} \right )^2
    \frac{1}{24}\left (\frac{k}{{\cal H}_{\rm c}}\right )^2,
\label{eq:cg}
\end{equation}
where ${\cal P}_h$ is the tensor power spectrum,
$\Omega_{r,0}h^2 = 4.2 \times 10^{-5}$ stands for the current radiation density
parameter for massless neutrinos \cite{PhysRevD.110.030001}, and the subscripts ``c'' and ``f'' denote
evaluation at the time $\eta_\uc \gg 1/k$, after which the GW emission of a
given mode $k$ has become negligible, and at a reference time
$\eta_{\rm f}$ at which the background has settled back to
$w = c_s^2 = 1/3$, respectively. The overline in Eq.~\eqref{eq:OmegaGW}
denotes a time average over multiple GW oscillations, inherited from the
definition of the GW energy density,
$\rho_\text{GW} \propto \overline{ \langle (h^{\prime})^2 \rangle}$~\cite{Maggiore:1999vm}. The prefactor
\begin{equation}
    \left (\frac{a_{\rm c}{\cal H}_{\rm c}}{a_{\rm f}{\cal H}_{\rm f}} \right )^2
    = \left ( \frac{g_{*\rho}}{g_{*\rho}^0} \right) \left( \frac{g_{*s}^0}{g_{*s}}\right)^{4/3},
\end{equation}
tracks the different dilution of the energy density in the GW sector compared
to the background plasma, which is particularly relevant across the QCD
phase. 
We have identified with the superscript $^0$ the present-day values. 
 On top of this, the smaller $c_s^2$ encountered around the QCD era
delays the oscillation of density perturbations after their Hubble re-entry
compared to a pure radiation background. This second effect is captured by
the transfer functions computed below. For concreteness, following the choice
of Ref.~\cite{Abe:2020sqb}, we fix $\eta_\uc =400/k$. This choice ensures GW
modes are evolved to the point at which they are sufficiently deep inside the
horizon to scale as a radiation component ($h\sim 1/a$); from then on, their
abundance relative to radiation is only sensitive to changes in the
relativistic degrees of freedom.

We adopt the Green's function method to solve for $h_{\vec k}(\eta)$,
\begin{align}
    a(\eta) h_{\vec k}(\eta)
    = 4 \int^\eta \d\bar{\eta}\, G_{\vec k}(\eta, \bar{\eta})\, a(\bar{\eta})\, {\cal S}_{\vec k}(\bar{\eta}),
\end{align}
where $G_{\vec k}(\eta, \bar{\eta})$ is the Green's function of
$\Lambda_\eta$ in Eq.~\eqref{eq:Lammbda_eta} and thus solves
\begin{align}
    G_{\vec k}''(\eta, \bar{\eta}) + \left( k^2 - \frac{ a''(\eta)}{a(\eta)}\right) G_{\vec k}(\eta, \bar{\eta}) = \delta (\eta - \bar{\eta}), \label{EOM_Green}
\end{align}
with
\begin{equation}
    \frac{a''(\eta)}{a(\eta)} = \frac{1-3w(\eta)}{2}\,{\cal H}^2(\eta) .
\label{eq:app_over_a}
\end{equation}
The two independent homogeneous solutions $g_{ik}(\eta)$, with
$i\in \{1,2\}$, entering the Green's function satisfy the tensor mode
equation
\begin{equation}
    g_{ik}''(\eta) + \left [k^2 - \frac{a''(\eta)}{a(\eta)} \right ] g_{ik}(\eta) = 0 ,
\label{eq:tensor_mode_eq}
\end{equation}
equivalent to $\Lambda_\eta\, g_{ik}(\eta) = 0$. The Green's function can be expressed as
\begin{equation}
    G_{\vec k}(\eta, \bar{\eta})
    = \frac{1}{{\cal N}_k}
     \left [ g_{1k}(\eta) g_{2k}(\bar \eta)- g_{1k}(\bar \eta) g_{2k}( \eta) \right ],
\end{equation}
where
${\cal N}_k=g_{1k}^\prime(\bar{\eta})g_{2k}(\bar{\eta})-g_{1k}(\bar{\eta})g_{2k}^\prime(\bar{\eta})$
is the (constant) Wronskian. We adopt the initial conditions
$g_{1k}(\eta_{\rm in})=0$, $g_{1k}'(\eta_{\rm in})=1$ and
$g_{2k}(\eta_{\rm in})=1$, $g_{2k}'(\eta_{\rm in})=0$, for which
${\cal N}_k=1$.

To perform the momentum integration, we introduce the loop variables
\begin{equation}
t \equiv u+v-1 \in[0,\infty),
\quad
s \equiv u-v \in[-1,1],
\label{eq:ts_def}
\end{equation}
where $u=|{\vec k}-{\vec q}|/k$ and $v= q/k$ are the internal momenta in
units of the external one, related to $(t,s)$ by $u=(t+s+1)/2$ and
$v=(t-s+1)/2$. The momentum-conservation triangle inequality is exactly the
half-strip $t\ge0$, $|s|\le1$, a $k$-independent and rectangular domain.

The kernel can then be expanded as a combination of the two tensor mode
functions multiplied by slowly varying factors,
\begin{align}
    I(k,t,s,\eta)\!= \!g_{1k}(\eta)I_2(k,t,s,\eta)\!-\!g_{2k}(\eta)I_1(k,t,s,\eta),
\label{eq:kernel_I}
\end{align}
with
\begin{align}
    I_i(k,t,s,\eta)
    =
    \frac{1}{a({\eta})}
    \frac{k^2}{{\cal N}_k}\int_{\eta_{\rm in}}^\eta\dd{\bar{\eta}}\,
    g_{ik}(\bar{\eta})\,
    a(\bar{\eta})\,
    f(t,s,\bar \eta) ,
\label{eq:Ii}
\end{align}
for $i=1,2$, where $\eta_\text{in}$ is the initial integration time, chosen to capture the whole  epoch in which GW emission was active. The function $f$ is the source of
Eq.~\eqref{eq: EoM_gw} expressed in terms of the transfer functions
$\Phi_1 \equiv \Phi_{uk}(\bar\eta)$ and $\Phi_2 \equiv \Phi_{vk}(\bar\eta)$
and normalized to the primordial curvature perturbation, fully accounting for a time-dependent equation of state (see, e.g., Ref.~\cite{Kohri:2018awv})
\bme{
f(t,s,\bar\eta) =
\frac{6(w+1)}{3w+5}\,\Phi_1\Phi_2
+\frac{3(1+3w)^2(1+w)}{(3w+5)^2}\,\frac{\Phi_1'\Phi_2'}{\mathcal H^2} \\
+\frac{6(1+3w)(w+1)}{(3w+5)^2}\,\frac{\Phi_1'\Phi_2+\Phi_2'\Phi_1}{\mathcal H},
\label{eq:source_f}
}
with $w$ and $\mathcal H$ evaluated at $\bar\eta$. In the radiation-dominated limit, we have ${\mathcal H}=1/\bar\eta$ and $w=1/3$, as well as the gravitational potential being described by an analytic oscillatory function. In our scenario, by contrast, each input of Eq.~\eqref{eq:source_f} is affected by the time-dependent equation of state: $w$ and $\mathcal H$ enter explicitly through the prefactors, while the evolutions of $\Phi_1$ and $\Phi_2$ must be obtained from a full numerical integration.

\begin{figure*}[t!]
\centering
\includegraphics[width=0.95\textwidth]{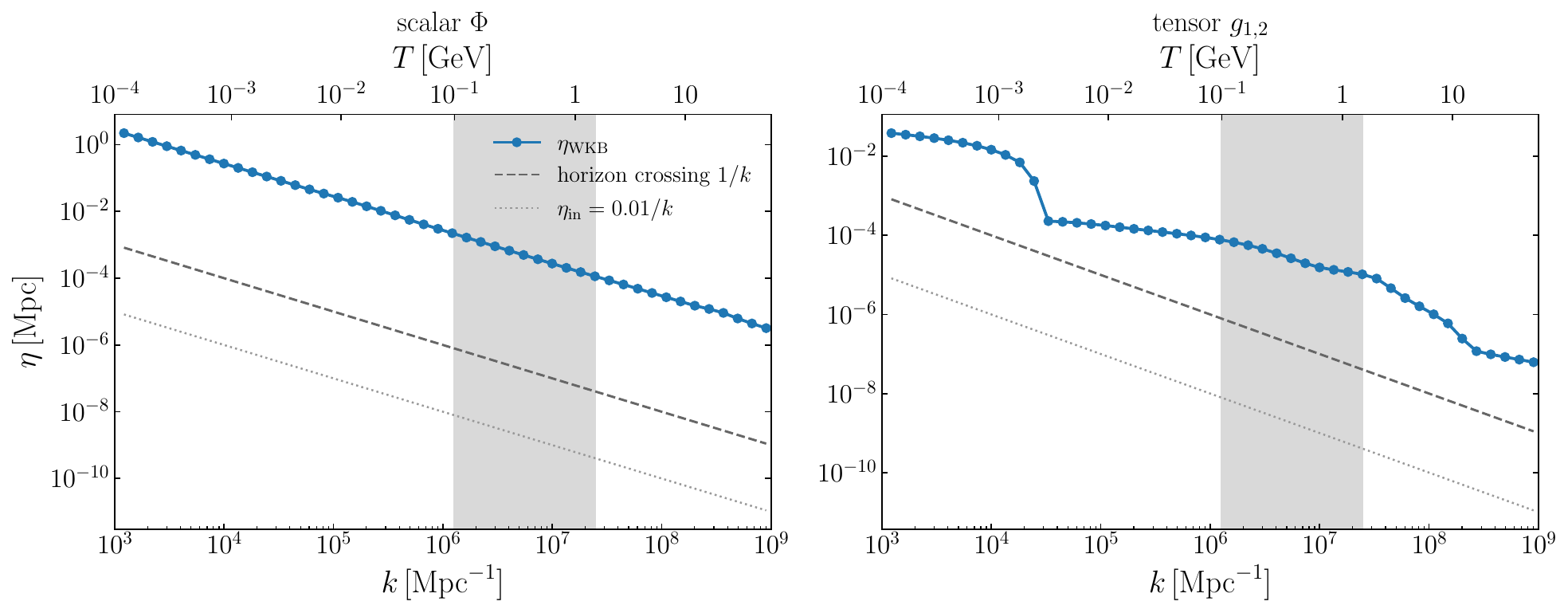}
\caption{Matching time $\eta_{\rm WKB}(k)$ beyond which the WKB form of
Eq.~\eqref{eq:wkbsol} is adopted, for the scalar transfer function (left) and
the tensor mode functions (right), compared with the horizon-crossing time
$1/k$ (dashed) and the initial time $\eta_{\rm in}=0.01/k$ (dotted). 
The top
axes report the plasma temperature at horizon crossing of each mode.}
\label{fig:etawkb}
\end{figure*}

As the scalar perturbations are efficiently damped after Hubble crossing, the
emission has converged by $\eta_\uc$, and the kernel subsequently oscillates
only through the mode functions $g_{1k}(\eta)$ and $g_{2k}(\eta)$. The time
average entering Eq.~\eqref{eq:OmegaGW} is taken over several oscillation
periods around the evaluation time $\eta_\uc$, where both $g_{ik}$ oscillate
freely with constant envelope. It therefore acts only on the products of
tensor mode functions,
\begin{align}
\overline{I(k,t_1,s_1) I(k,t_2,s_2)} 
& \simeq
I_2(k,t_1,s_1) I_2(k,t_2,s_2)\,
\overline{g_{1k}^2} 
\nonumber \\
&- \left [I_1(k,t_1,s_1)I_2(k,t_2,s_2)
\right.  \nonumber 
\\
 & \left. + I_2(k,t_1,s_1)I_1(k,t_2,s_2) \right ]\, \overline{g_{1k}g_{2k}}
 \nonumber
\\
&    + I_1(k,t_1,s_1) I_1(k,t_2,s_2)\,
    \overline{g_{2k}^2} ,
\label{eq:osc_avg}
\end{align}
where deep inside the horizon $g_{1k}$ and $g_{2k}$ oscillate rapidly.
In RD, they do so in quadrature and the cross term $\overline{g_{1k}g_{2k}}$ averages to zero over an
integer number of periods, while
$\overline{g_{1k}^2}$ and $\overline{g_{2k}^2}$ reduce to half the squared
envelopes. However, this simplification does not hold with a time-varying equation of state, and thus each piece should be retained. It is therefore necessary to compute $I_1(k,t,s)$ and $I_2(k,t,s)$,
together with the averaged product of tensor mode functions, all evaluated at
$\eta_\uc(k) = 400/k$.

At leading order in the curvature perturbation, the oscillation-averaged
tensor power spectrum reads \cite{Espinosa:2018eve,Kohri:2018awv}
\begin{align}
    \overline{{\cal P}_h^\text{\tiny G} (k)} =
    \frac{1}{4}
    \int_0^\infty \d t
    \int_{0}^{1}\dd{s}
    \frac{\overline{J^2(k,t,s)} }{(u v)^2}
     {\cal P}_\zeta \left ( k u \right )
     {\cal P}_\zeta \left ( k v \right ) ,
 \label{eq:P_h_ts}
\end{align}
where ${\cal P}_\zeta$ is the dimensionless power spectrum of the curvature
perturbation and we introduced~\cite{Li:2023qua}
\begin{align}
     J(k,t,s) \equiv t (t+2) (1-s^2)\, I(k,t,s) ,
\label{eq:Jdef}
\end{align}
to make the notation more compact. The integrand in this case is even in $s$, which is
used to restrict the integral to $s \ge 0$ at the price of an extra factor
of two relative to Ref.~\cite{Kohri:2018awv}.

\subsection{Sub-horizon evolution: WKB continuation}\label{sec:wkb}

Tracking the sub-horizon evolution numerically is very expensive and more susceptible to inaccuracies, particularly in the squeezed limit for which $q \gg k$. Therefore, we adopt a numerical approach based on the WKB approximation. 
Both Eq.~\eqref{eq:phievo} and the homogeneous tensor equation take the form
\begin{equation}
X'' + f_1\,X' + \left (k^2 f_3+f_2 \right )X=0 ,
\label{eq:unified}
\end{equation}
with
\begin{equation}
   \left  \{f_1,f_2,f_3 \right \}=\left (3\mathcal H(1+c_s^2),\,3\mathcal H^2(c_s^2-w),\,c_s^2 \right ),
\end{equation}
for the scalar $\Phi$ and
\begin{equation}
\left  \{f_1,f_2,f_3 \right \}=\left (0,\,-\tfrac{1-3w}{2}\mathcal H^2,\,1 \right ),
\end{equation} 
for the tensor
mode functions $g_{1,2}$. The substitution
$X=\psi\,\exp\left(-\tfrac12\int f_1\,\d\eta\right)$ removes the friction
term exactly, yielding $\psi''+\omega^2\psi=0$ with
\begin{equation}
\omega^2 = k^2f_3 + f_2 - \frac{1}{2}f_1' - \frac{1}{4}f_1^2 .
\label{eq:omega2}
\end{equation}
Deep inside the horizon, the frequency is dominated by the first term and
varies adiabatically, so the solution is well approximated by the WKB form 
\begin{align}
X(\eta) &\simeq e^{-y(\eta)/2}\,f_3^{-1/4}
\left [C_+e^{\,ikz(\eta)}+C_-e^{-ikz(\eta)} \right ],
\nonumber \\
y(\eta)&=\int_{\eta_{\rm WKB}}^{\eta} f_1\,\d\eta',
\quad
z(\eta)=\int_{\eta_{\rm WKB}}^{\eta}\sqrt{f_3}\,\d\eta' ,
\label{eq:wkbsol}
\end{align}
with the constants $C_\pm$ fixed by matching $X$ and $X'$ to the numerical
solution at the matching time $\eta_{\rm WKB}$. The latter is defined as the
earliest time beyond which
\begin{equation}
|f_2| + \frac{1}{2}|f_1'| + \frac{1}{4} f_1^2 \;<\; \epsilon\,k^2 f_3
\label{eq:wkbcrit}
\end{equation}
holds for the remainder of the evolution, with $\epsilon=10^{-3}$. We have also checked the standard adiabatic condition $|\omega'/\omega^2|\ll1$ directly. For the modes considered in this study, this condition is satisfied well before the conservative criterion in Eq.~\eqref{eq:wkbcrit}. The
functions $y$ and $z$ are smooth, so no oscillation needs to be resolved past
$\eta_{\rm WKB}$; this is what renders strongly hierarchical configurations,
in which the internal modes complete up to $10^{4}$--$10^{5}$ oscillations
before $\eta_\uc$, numerically tractable. Displacing $\eta_{\rm WKB}$ for the
tensor modes by a full order of magnitude changes $\Omega_{\rm GW}$ by less
than $1\%$ across the nHz band.

Figure~\ref{fig:etawkb} shows $\eta_{\rm WKB}(k)$ for the scalar and tensor
modes across the wavenumbers relevant for the QCD band. For the scalar, the
matching occurs after several sub-horizon oscillations,
$k\eta_{\rm WKB}=\mathcal O(10^2$--$10^4)$; for the tensor modes it occurs
shortly after horizon crossing, with a visible feature where the crossover
modifies $a''/a$ and delays the saturation of the criterion in
Eq.~\eqref{eq:wkbcrit}. 
In the latter case, as in RD, the WKB ansatz is an exact solution of the equations of motion; $\eta_{\rm WKB}$ can approach horizon crossing, as the strong hierarchy $k\eta_{\rm WKB} \gg 1$ is not needed as long as $w \to 1/3$, as in the intermediate scales around ${\rm few} \times 10^4\,{\rm Mpc}^{-1}$.
The complete list of technical choices is collected
in Table~\ref{tab:setup}. 

\subsection{Kernel time integration}\label{sec:timeint}

The integrals in Eq.~\eqref{eq:Ii} involve the product of three oscillatory
functions: the two scalar legs, oscillating with instantaneous frequencies
$c_s u k$ and $c_s v k$, and the tensor mode function at frequency $k$. The
integrand is evaluated on a logarithmic grid in $\bar\eta$ dense enough to
resolve the fastest internal frequency present, with the grid onset set by
the internal momenta, $\bar\eta_{\rm in} = 10^{-2}/\max(u,v,1)k$. We have
verified the convergence of this quadrature against high-resolution
references, both on the QCD background and in exact RD,
where every ingredient of the integrand is known in closed form.

\subsection{Tabulated kernels}\label{sec:tables}

The oscillation-averaged kernel $\overline{I^2(k,t,s)}$ is independent of the
primordial spectrum. We therefore tabulate
$(k\eta_\uc)^2\,\overline{I^2}$ on a fixed $(t,s)$ grid at $110$ values of the
external wavenumber at $\eta_\uc(k)=400/k$, log-spaced across $f\in[10^{-3},10^{4}]\,$nHz, with
$t\in[10^{-3},3\times10^{3}]$ ($1444$ points, clustered around the kernel
resonance) and $s\in[0,1)$ ($24$ points). Once the tables are available, the
spectrum induced by an arbitrary primordial spectrum follows from the
two-dimensional quadrature of Eq.~\eqref{eq:P_h_ts} over the stored grid, at
negligible computational cost: no mode function needs to be solved again.

The fixed $(t,s)$ window sets the hierarchy of scales a table node can
serve. The internal momenta satisfy $uk,vk \le (t_{\rm max}+2)k/2 \simeq
1.5\times10^{3}\,k$, so a node at wavenumber $k$ supports primordial spectra
peaked at $k_\ast$ up to
\begin{equation}
\frac{k}{k_\ast} \;\gtrsim \; \frac{2}{t_{\rm max}} \sim 10^{-3} .
\label{eq:hierarchy}
\end{equation}

\begin{table}[t]
\centering
\caption{Technical setup of the computation.} 
\label{tab:setup}
\begin{ruledtabular}
\begin{tabular}{ll}
initial time & $\eta_{\rm in} =  10^{-2}/\max(u,v,1)k$ for each mode\\
& (external $k$; internal $uk$, $vk$)\\
sub-horizon evolution & numerical integration up to $\eta_{\rm WKB}$,\\
& WKB continuation beyond [Eq.~\eqref{eq:wkbcrit}]\\
evaluation time & $\eta_\uc = 400/k$, i.e.\ $x_\uc \equiv k\eta_\uc = 400$~\cite{Abe:2020sqb}\\
oscillation average & last $8$ periods of the external mode\\
& before $\eta_c$ [Eqs.~\eqref{eq:OmegaGW} and \eqref{eq:osc_avg}]\\
tabulated kernel & $t\in[10^{-3},3\times10^{3}]$ ($1444$ points),\\
& $s\in[0,1)$ ($24$ points),\\
& $110$ nodes in $k$, $f\in[10^{-3},10^{4}]\,$nHz\\
\end{tabular}
\end{ruledtabular}
\end{table}

\begin{figure}[t!]
\centering
\includegraphics[width=0.48\textwidth]{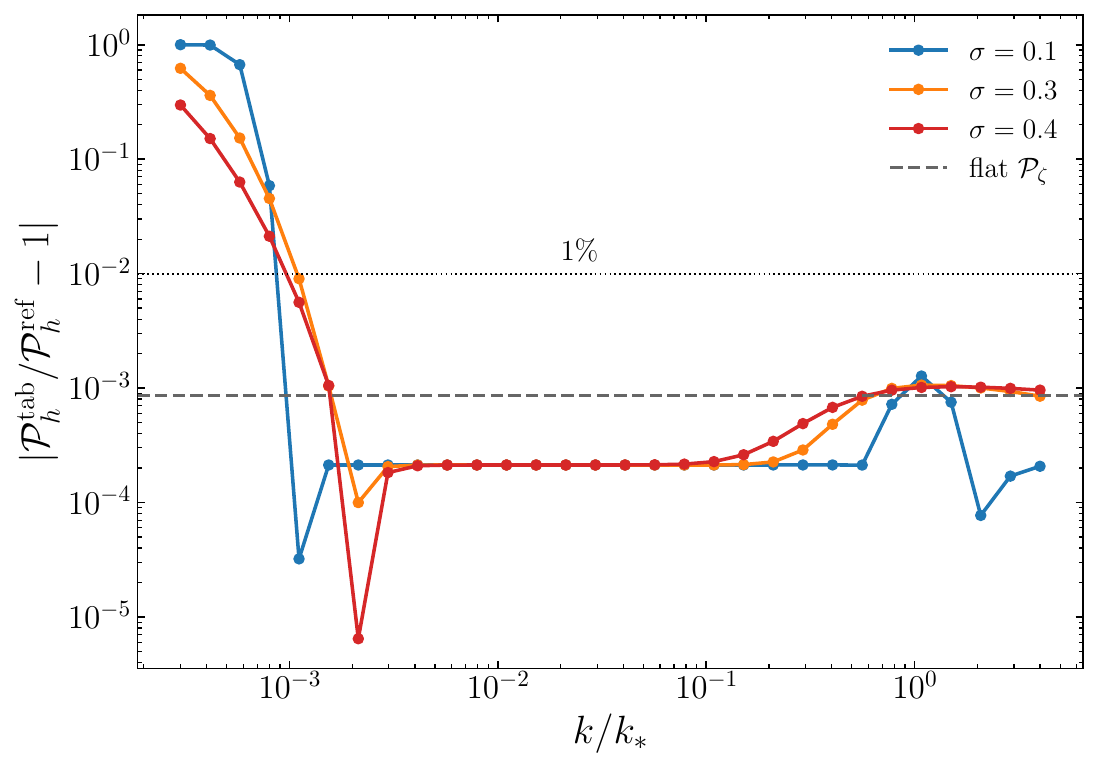}
\caption{Accuracy of the tabulated kernel representation as a function of
the hierarchy between the spectral peak $k_\ast$ and the table node $k$, for
lognormal spectra of width $\sigma$ and for a scale-invariant spectrum
(dashed) evaluated for the radiation-dominated universe. The plateau marks the intrinsic discretization accuracy of the
$(t,s)$ grid; the rise at $k/k_\ast\sim10^{-3}$ reflects the finite extent of
the tabulated window, Eq.~\eqref{eq:hierarchy}.}
\label{fig:coverage}
\end{figure}

Figure~\ref{fig:coverage} quantifies the accuracy of the tabulated
representation: evaluating the exact RD kernel on the
table grid and comparing with a dense reference quadrature, the relative
error sits at the intrinsic discretization floor of a few$\,\times10^{-4}$
for lognormal spectra over the entire range $2\times 10^{-3}  \lesssim k/k_\ast \lesssim
4$. For lower values of $k/k_\ast$ ($k/k_\ast\lesssim10^{-3}$), we find the accuracy 
crossing the percent level, at which point the table should stop being
used. For a scale-invariant spectrum, which
engages the whole window at every node, the accuracy is
$9\times10^{-4}$. For the most hierarchical configurations stored in the
table, the time quadrature of Sec.~\ref{sec:timeint} contributes an
additional uncertainty below the percent level.

The full pipeline was also validated against the analytical result available in
exact RD~\cite{Kohri:2018awv}: for a scale-invariant
spectrum, the benchmark value $x_\uc^2\,\overline{{\cal P}_h}/A_\zeta^2 =
19.7$ is reproduced at the few-per-mille level, where $A_\zeta$ denotes the
amplitude of the curvature spectrum.

\section{Results}\label{sec:results}

We show the impact of QCD kernels on the shape of the SIGW spectrum by considering two limiting cases of a broad (here taken to be scale-invariant) and narrow (here approximated with a variable width lognormal) curvature power spectrum.

\subsection{Scale-invariant spectrum}

\begin{figure}[t!]
\centering
\includegraphics[width=0.48\textwidth]{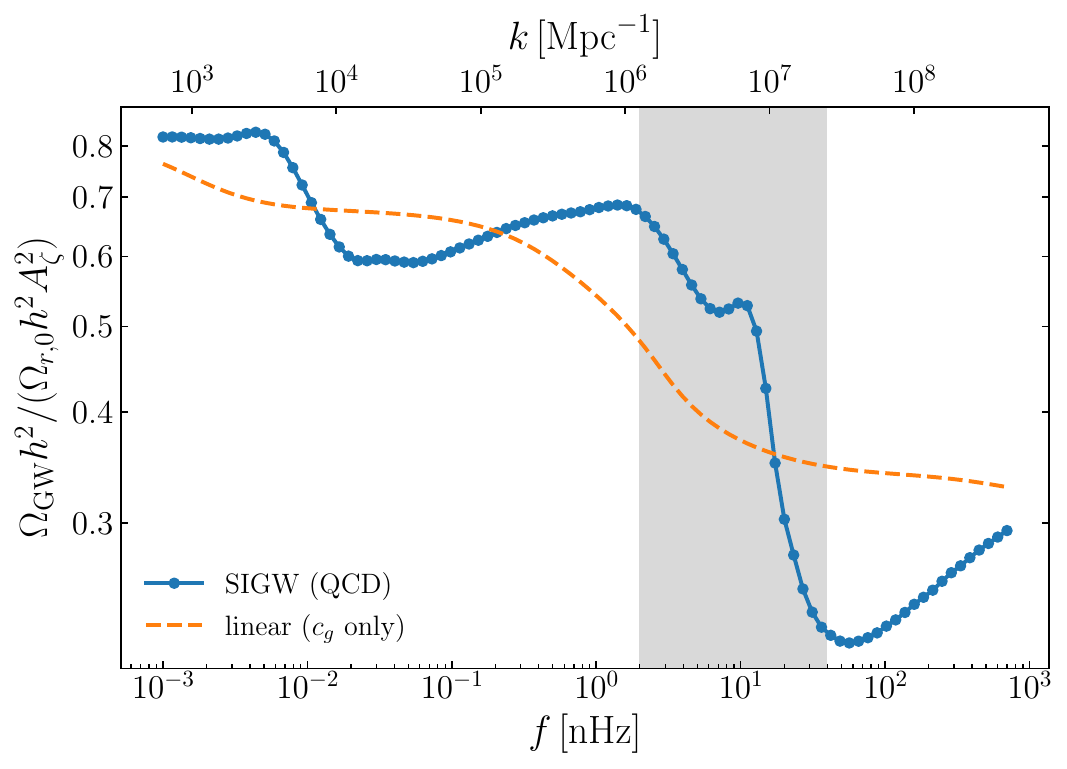}
\caption{Induced spectrum for a scale-invariant curvature spectrum (solid),
against the \emph{linear} reference combining the RD
kernel with the dilution factor $c_g(k)$ (dashed). The full result is
suppressed with respect to the reference above the QCD band and enhanced
below it, with the same pattern repeating, with smaller amplitude, around
the $e^-e^+$ annihilation scale.} 
\label{fig:flat}
\end{figure}

Fig.~\ref{fig:flat} shows the induced spectrum for a scale-invariant
curvature spectrum, ${\cal P}_\zeta = A_\zeta$, together with the
\emph{linear} reference obtained by combining the exact
RD kernel with the standard dilution factor included in $c_g(k)$ in
Eq.~\eqref{eq:cg}. The latter carries the entire effect of the modified
expansion history on the propagation; the difference between the two curves
therefore isolates the imprint of the crossover on the emission itself.

\begin{figure*}[t!]
\centering
\includegraphics[width=\textwidth]{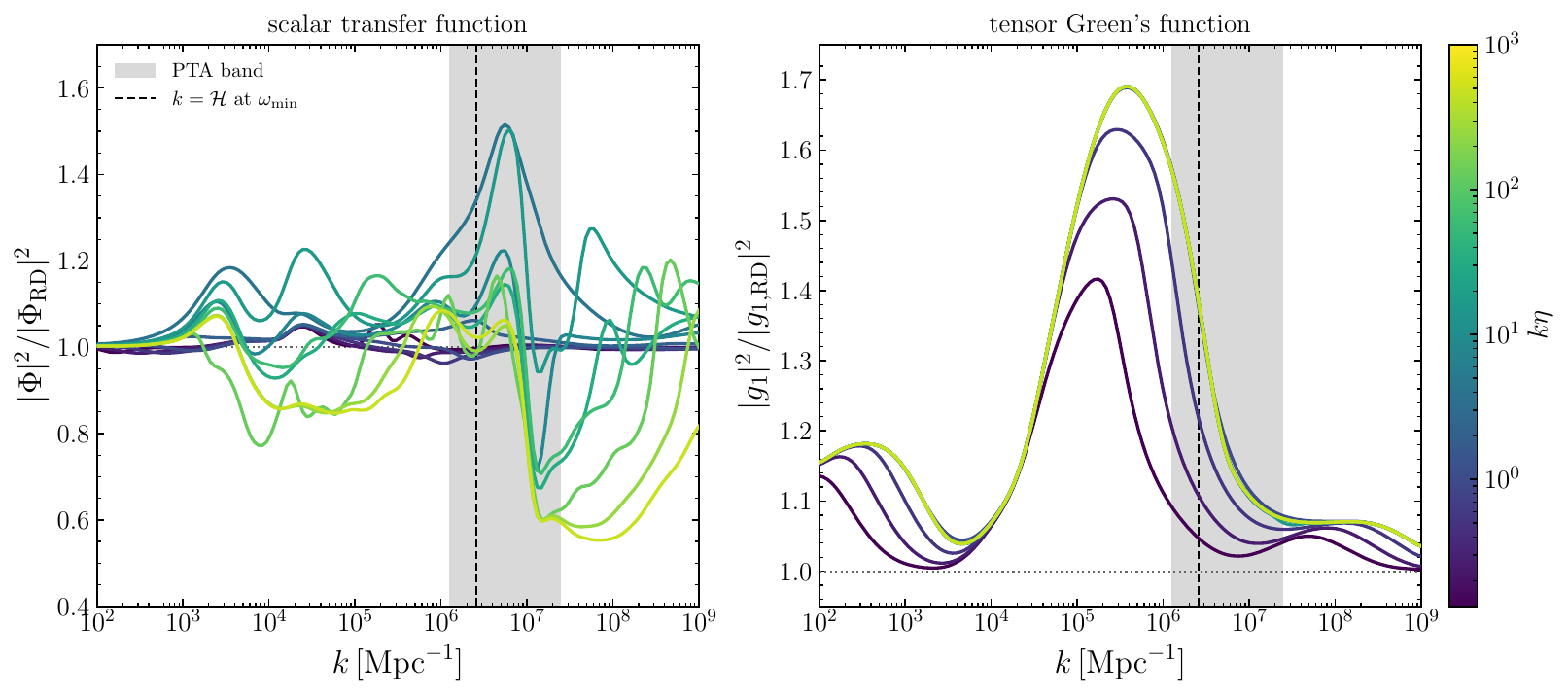}
\caption{Ratio of the scalar transfer function (left) and the tensor
mode function $g_1$ (right) on the QCD background to their
RD counterparts, as a function of the external wavenumber $k$,
for a few values of $k\eta$ spanning $[0.1,10^{3}]$ (color bar). 
Both panels show a temporary boost for modes whose relevant epoch overlaps a softened,
closer-to-matter equation of state (the main feature near $10^{6}\,{\rm
Mpc}^{-1}$ is the QCD crossover; the smaller one near $10^{8}\,{\rm
Mpc}^{-1}$ the approach to the electroweak scale); only the scalar
transfer function develops a dip below the RD value for
$k$ somewhat above each feature, a resonant effect tied to the sign change
of $c_s^2-w$ across the transition, absent for the tensor mode
functions.}
\label{fig:phigratio}
\end{figure*}

The ratio of the two curves displays a characteristic two-sided pattern
around each reduction of $w$ and $c_s^2$: the full result is suppressed, by up
to $\approx 35\%$, for modes crossing the horizon during the onset of the QCD era
($f \approx 50\,$nHz), and enhanced, by up to $\approx 35\%$, at
frequencies just below the band ($f \approx 2\,$nHz); a similar pattern appears around the $e^-e^+$ annihilation scale. Far
from each feature, the two curves coincide, as they must once every relevant
mode evolves entirely within an epoch with $w=c_s^2=1/3$.

The physical origin of the two-sided pattern traces back directly to the
scalar transfer function. 
Figure~\ref{fig:phigratio} shows
$|\Phi|^2/|\Phi_{\rm RD}|^2$ (left) and the analogous
tensor-mode-function ratio $|g_1|^2/|g_{1,\rm RD}|^2$ (right) on the QCD
background relative to RD, each line following a fixed
stage of sub/super-horizon evolution (fixed $k\eta$) as a function of $k$.
The friction term $\exp\!\left(-\frac12\int f_1\,\d\eta\right)$ of Sec.~\ref{sec:wkb}
only depends on $f_1=3\mathcal H(1+c_s^2)$. Softening the equation of state
increases the friction coefficient: $\mathcal H$ runs above its would-be RD value by enough to more than offset the smaller $c_s^2$.
However, the effect of this drag is modest and essentially the same for every mode that samples a given transition, contributing only a mild, systematic suppression. 
The
dominant, $k$-dependent boost-or-dip pattern instead originates in the
potential term $f_2$. For the scalar, $f_2=3\mathcal H^2(c_s^2-w)$ vanishes identically in radiation
domination but is nonzero and changes sign across each transition, since
$c_s^2$ and $w$ do not track each other exactly: a slowly varying
$f_2$ shifts the effective frequency adiabatically, and since the
oscillator's action $A^2\omega$ is then conserved, a locally positive
$f_2$ (frequency up) comes at the expense of a shrinking $\Phi$ envelope,
while a locally negative $f_2$ (frequency down) lets it grow. Since the
mode oscillates through the transition, the effect depends on the phase at which the mode encounters the sign change. The scalar ratio of Fig.~\ref{fig:phigratio} shows exactly a
temporary boost followed by a dip around each phase transition. The
tensor mode functions instead only ever show the boost: their potential
term, $f_2=-\frac{1-3w}{2}\mathcal H^2$
[Eq.~\eqref{eq:unified}], depends on $w$ alone, which never exceeds
$1/3$ during a softening episode, so $f_2$ keeps a fixed sign throughout.
The adiabatic response (lower effective frequency, larger
envelope) leads to an effective enhancement.
Through the quadratic source, the scalar boost is what feeds the
enhancement of the induced spectrum at $f\approx2\,$nHz in
Fig.~\ref{fig:flat}, and the scalar-only dip is the origin of the
suppression at $f\approx50\,$nHz there.


\begin{figure*}[t!]
\centering
\begin{subfigure}{0.47\textwidth}
\centering
\includegraphics[width=\textwidth]{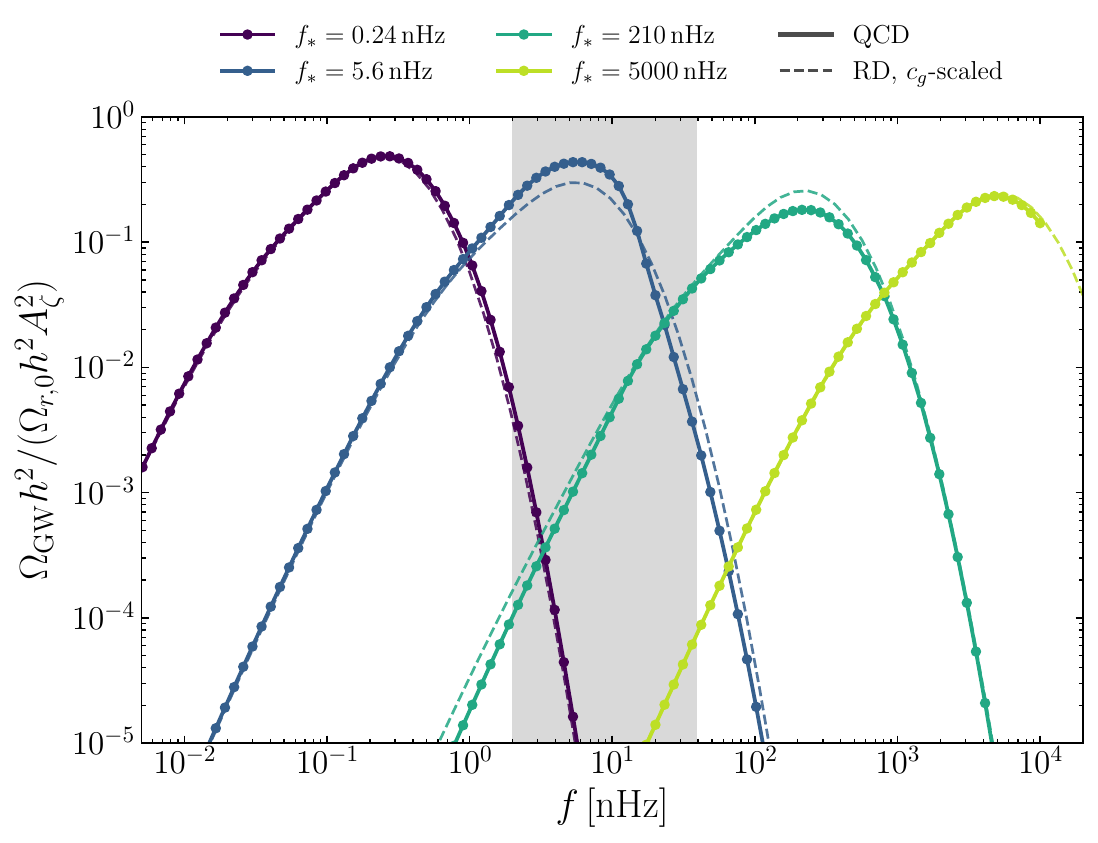}
\label{fig:lognormal}
\end{subfigure}
\hfill
\begin{subfigure}{0.52\textwidth}
\centering
\includegraphics[width=\textwidth]{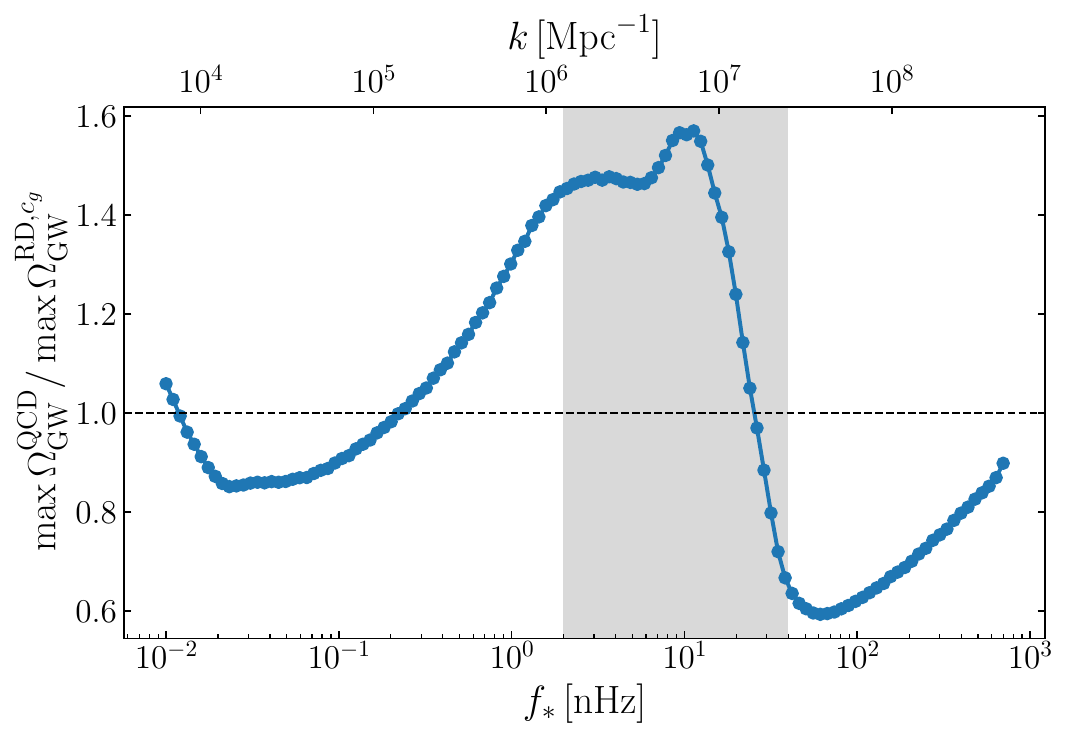}
\label{fig:peakratio}
\end{subfigure}
\caption{\emph{Left}: induced spectra for lognormal curvature spectra of width
$\sigma=0.4$ and different peak frequencies $f_\ast$ (solid, computed on the
QCD background from the tabulated kernels), compared with the
RD kernel rescaled by the dilution factor $c_g$
(dashed). \emph{Right}: ratio between the peak amplitude of the induced spectrum
computed on the QCD background and that of the $c_g$-rescaled
RD prediction, as a function of the peak frequency
$f_\ast$ of the lognormal curvature spectrum.}
\label{fig:lognormal_and_peakratio}
\end{figure*}

\subsection{Peaked spectra}

Figure~\ref{fig:lognormal_and_peakratio} (left) shows the induced spectra
for lognormal curvature spectra,
\begin{equation}
{\cal P}_\zeta(k)=\frac{A_\zeta}{\sqrt{2\pi\sigma^2}}
\exp\left [-\frac{\log_{10}^2 (k/k_\ast)}{2\sigma^2} \right ],
\label{eq:lognormal}
\end{equation}
of width $\sigma$ and several peak frequencies $f_\ast$ spanning the PTA
band, compared with the RD prediction rescaled by the same
dilution factor $c_g$. We fix $\sigma$ to the representative value $0.4$
throughout, corresponding to a relatively narrow spectrum of finite
physical width. Figure~\ref{fig:lognormal_and_peakratio} (right)
summarizes the effect on the peak amplitude by showing the ratio of the peak heights of
the two calculations, as a function of $f_\ast$. The peak is enhanced by up to $\approx 40\%$ for
$f_\ast \approx (2\text{--}10)\,\si{nHz}$ and $\approx 55\%$ for $f_\ast \approx (10\text{--}20)\,\si{nHz}$. At higher peak frequencies,
$f_\ast \approx (30\text{--}100)\,\si{nHz}$, we see a suppression of up to
$\approx 40\%$, with a
milder replica of the same modulation at the $e^-e^+$ scale. Both features
directly track the two-sided pattern discussed in the previous subsection, now weighted by the
narrow spectral support around $k_\ast$.

\begin{figure*}[t!]
\centering
\includegraphics[width=0.49\textwidth]{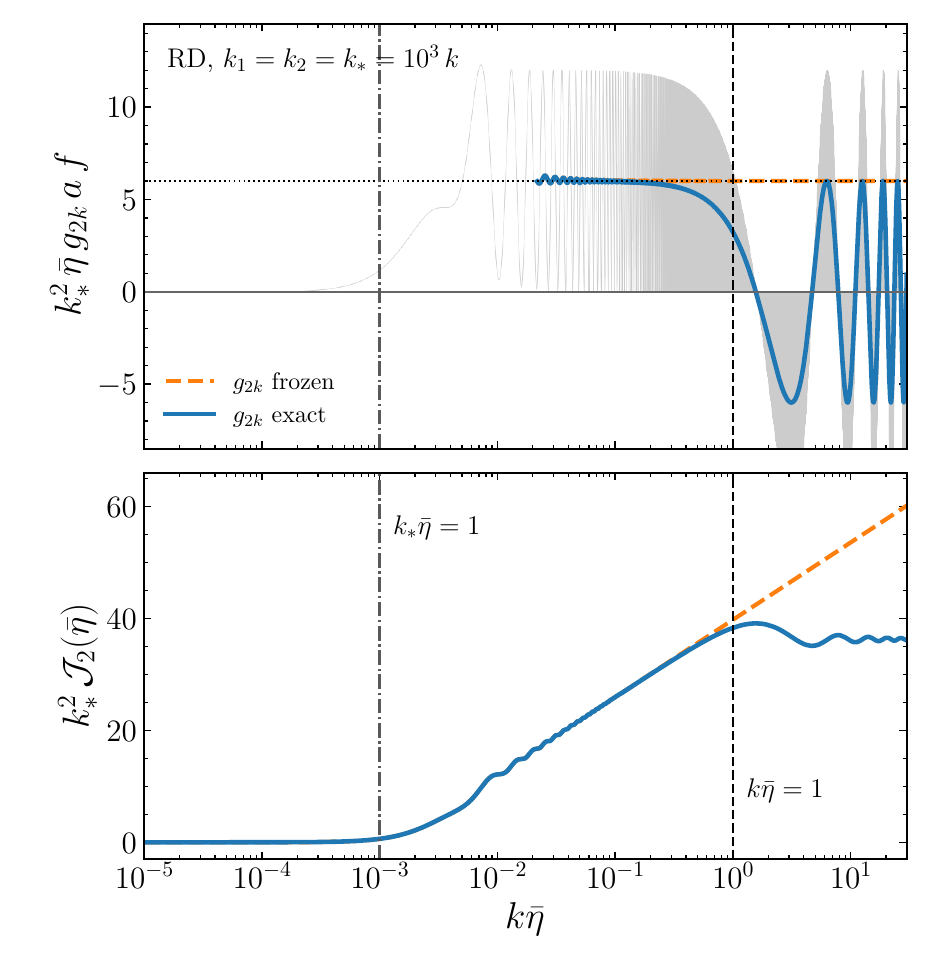}\hfill
\includegraphics[width=0.49\textwidth]{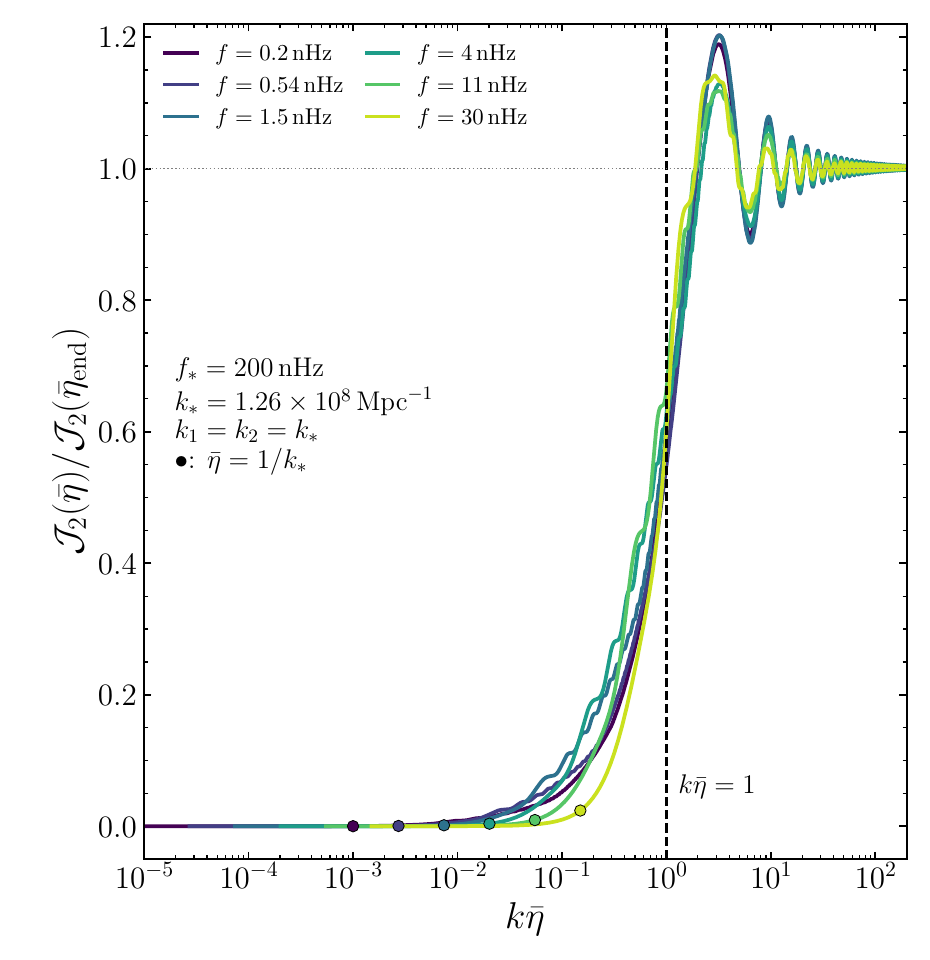}
\caption{Saturation of the running source integral $\mathcal J_2$ of
Eq.~\eqref{eq:J2running} at horizon entry of the observed mode, in the
squeezed configuration $k_1=k_2=k_\ast$ ($s=0$, $t=2k_\ast/k-1$). \emph{Left}: exact RD with $k_\ast=10^3k$.
\emph{Top}: contribution per $e$-fold, $k_\ast^2\,\bar\eta\,g_{2k}\,a\,f$ (gray:
raw; blue: averaged over the fast source oscillation; orange: $g_{2k}$ kept
frozen at unity); the dotted line marks the analytic plateau. 
\emph{Bottom}:
$k_\ast^2\mathcal J_2$, growing as $\ln(k_\ast\bar\eta)$ between the two
horizon entries and saturating after $k\bar\eta\simeq1$.
\emph{Right}: $\mathcal J_2(\bar\eta)/\mathcal J_2(\bar\eta_{\rm end})$ on the
QCD background for $f_\ast=200\,\si{nHz}$ and six observed frequencies
$f\in[0.2,30]\,\si{nHz}$ on the IR tail; dots mark $\bar\eta=1/k_\ast$, the dashed line
$k\bar\eta=1$. We indicate with $\bar \eta_{\rm end}$ a late time at which emission becomes negligible. }
\label{fig:irtail}
\end{figure*}

Turning now to the infrared tail, $f \ll f_\ast$, the QCD effect is qualitatively different from the peak rescaling of Fig.~\ref{fig:lognormal_and_peakratio}. Specifically, the infrared tail of the induced spectrum
does not follow the universal causality scaling $\Omega_{\rm GW}\propto k^3$, nor the predicted universal modulation induced by QCD \cite{Franciolini:2023wjm}.
That scaling holds for sources bounded both in frequency and in time, acting
during RD, and probed at wavenumbers below every physical
scale of the source, including its duration~\cite{Cai:2019cdl}. The
scalar-induced source is not bounded in time. Once the peak-scale modes are
inside the horizon, $\Phi_i'\propto 1/(k_\ast\bar\eta^2)$ while
$a\propto\bar\eta$
, so the combination entering Eq.~\eqref{eq:Ii} scales as
$a(\bar\eta)f(t,s,\bar\eta)\propto 1/\bar\eta$, with fast oscillations at the sound frequency of the peak-scale modes, plus
the Green's function.  As long
as the observed tensor mode is still super-horizon, $g_{1k}(\bar\eta)\simeq0$ and $g_{2k}(\bar\eta)\simeq1$, so
every $e$-fold of $\bar\eta$ contributes equally to the running integral
\begin{equation}
\mathcal J_2(\bar\eta)\equiv\int_{\eta_{\rm in}}^{\bar\eta}
\dd{\bar\eta'}g_{2k}(\bar\eta')\,a(\bar\eta')\,f(t,s,\bar\eta') ,
\label{eq:J2running}
\end{equation}
in terms of which $I_2(k,t,s,\eta)=k^2\mathcal J_2(\eta)/a(\eta)$, so that
$\mathcal J_2$ grows as $\ln(k_\ast\bar\eta)$. The growth stops only when the
observed mode itself enters the horizon, $k\bar\eta\simeq1$: from then on
$g_{2k}$ oscillates at frequency $k$ against a source envelope that varies on
the Hubble time, and successive contributions cancel. The effective duration
of the emission thus grows with decreasing $k$, the kernel acquires a factor
$\ln(k_\ast/k)$, and in RD the tail of a narrow spectrum
becomes $\Omega_{\rm GW}\propto k^3\ln^2(k_\ast/k)$, i.e.\ the log-dependent
slope $n_{\rm GW}=3-2/\ln(f_\ast/f)$ of Ref.~\cite{Yuan:2019wwo,LISACosmologyWorkingGroup:2024hsc}.
Figure~\ref{fig:irtail} (left) isolates this mechanism in exact radiation
domination for $k_1=k_2=k_\ast=10^3k$: $k_\ast^2\mathcal J_2$ reaches $38.3$
at $k\bar\eta=1$, close to the pure-logarithm value $6\ln 10^3\simeq41.4$,
and then saturates, whereas keeping $g_{2k}$ frozen would make it grow
indefinitely. 

Whenever the emission window overlaps with a softening of the equation of state, this exact template is modified.  
Figure~\ref{fig:irtail} (right) shows the corresponding behavior on the
QCD background, for $f_\ast=200\,\si{nHz}$ and six observed frequencies
between $0.2$ and $30\,\si{nHz}$: the accumulation starts at
$\bar\eta\simeq1/k_\ast$ and freezes at $k\bar\eta\simeq1$ for every $k$.
For observed frequencies below a few $\si{nHz}$, this window contains the
crossover, so the tail integrates the thermal history over the whole interval
$\bar\eta\in[1/k_\ast,1/k]$, weighted at each epoch by the local $w$,
$c_s^2$ and $\mathcal H$, rather than sampling a single epoch. 
%
%
Consequently, the overall normalization of the tail is set by $c_g$ evaluated at the peak frequency $f_*$ rather than at $f$, because the tensor mode is continuously replenished and is not subject to the relative suppression that entropy injection into the non-GW sector imprints on modes that free stream.
The tail on the QCD background therefore follows neither the causality scaling nor its RD logarithmic template (see also \cite{Yuan:2026krm}). 

In summary, the QCD crossover leaves an imprint on the induced background
that is not a small correction: across the frequency window
$f \in [1/T_{\rm obs}, 20/T_{\rm obs}]$ probed by a $16$-year PTA dataset, shaded in Figs.~\ref{fig:eos}, \ref{fig:etawkb},
\ref{fig:flat}, \ref{fig:phigratio}, and~\ref{fig:lognormal_and_peakratio}, the height of the
induced spectrum is modified by up to $\approx 55\%$ with respect to the
RD expectation, with both signs realized depending on
whether the dominant emission occurs before or after the transition. 
The modification is not confined to the peak: the infrared tail departs from both the causality scaling and its RD logarithmic template.
Therefore, the QCD crossover is a SM effect that cannot be neglected in the interpretation of a putative scalar-induced origin of the
observed signal, nor in the inference of the underlying curvature power
spectrum and the associated PBH abundance.

\section{Relevance for PTA experiments}
\label{sec:pta}

The size of the effect discussed above motivates asking whether it significantly shifts
the curvature-spectrum parameters inferred from actual PTA data (see, e.g., Ref.~\cite{Abe:2023yrw} for an earlier work). Following Ref.~\cite{Ellis:2023oxs}, we fit the NANOGrav 15-year free-spectrum posterior
\cite{NANOGrav:2023gor}, given as $14$ tabulated one-dimensional posteriors
in $\Omega_{\rm GW}$, one per frequency bin $f_n = n/T_{\rm obs}$ with
$T_{\rm obs}=16.03\,$yr, with a simplified broken-power-law curvature spectrum
\begin{equation}
{\cal P}_\zeta(k) = A\,
\frac{\alpha+\beta}
{\beta\,(k/k_\ast)^{-\alpha}
      +\alpha\,(k/k_\ast)^{\beta}} ,
\label{eq:bpl}
\end{equation}
so that ${\cal P}_\zeta= A$ at $k=k_\ast$, ${\cal P}_\zeta\propto k^{\alpha}$ far below the
peak scale and ${\cal P}_\zeta\propto k^{-\beta}$ far above it. 

With only $14$ noisy bins on the rising infrared part of the spectrum,
both $\alpha$ and $\beta$ remain essentially prior-dominated: $\beta$,
which sets the falling tail $k\gg k_\ast$, is unconstrained because the
NANOGrav band only resolves the infrared rise and, at best, the onset of
the peak, never its high-$k$ side; $\alpha$, which sets the infrared tail
$k\ll k_\ast$, is likewise unconstrained for
$\alpha\gtrsim2$, since within the resolved band the spectrum is then
already steep enough to be degenerate with any yet steeper,
causality-limited tail \cite{Byrnes:2018txb,Cole:2022xqc}, so the data cannot distinguish among the allowed
values above this threshold.
We
sample $\log_{10}A$, $\log_{10}k_\ast$, $\alpha$ and $\beta$ with flat
priors, once with the exact-RD kernel and once with the QCD-crossover
kernel of this work, keeping every other assumption identical, so that
any shift between the resulting posteriors isolates the effect of the
QCD crossover.

Typically associated with the broken-power-law power spectrum described by
Eq.~\eqref{eq:bpl} is a sharpness parameter, which controls how sharply the spectrum transitions between the two power-law asymptotic behaviors. We choose to hold this fixed to unity for a smooth transition in both fits rather than
sample it --- with only $14$ noisy NANOGrav bins on the rising infrared part
of the spectrum, the data constrain the local slope and curvature of
${\cal P}_\zeta$ around $k_\ast$ but not independently how sharply the two
power-law branches join, so a free transition parameter remains unconstrained. It is therefore not included in the analysis.

\begin{figure*}[t!]
\centering
\includegraphics[width=\textwidth]{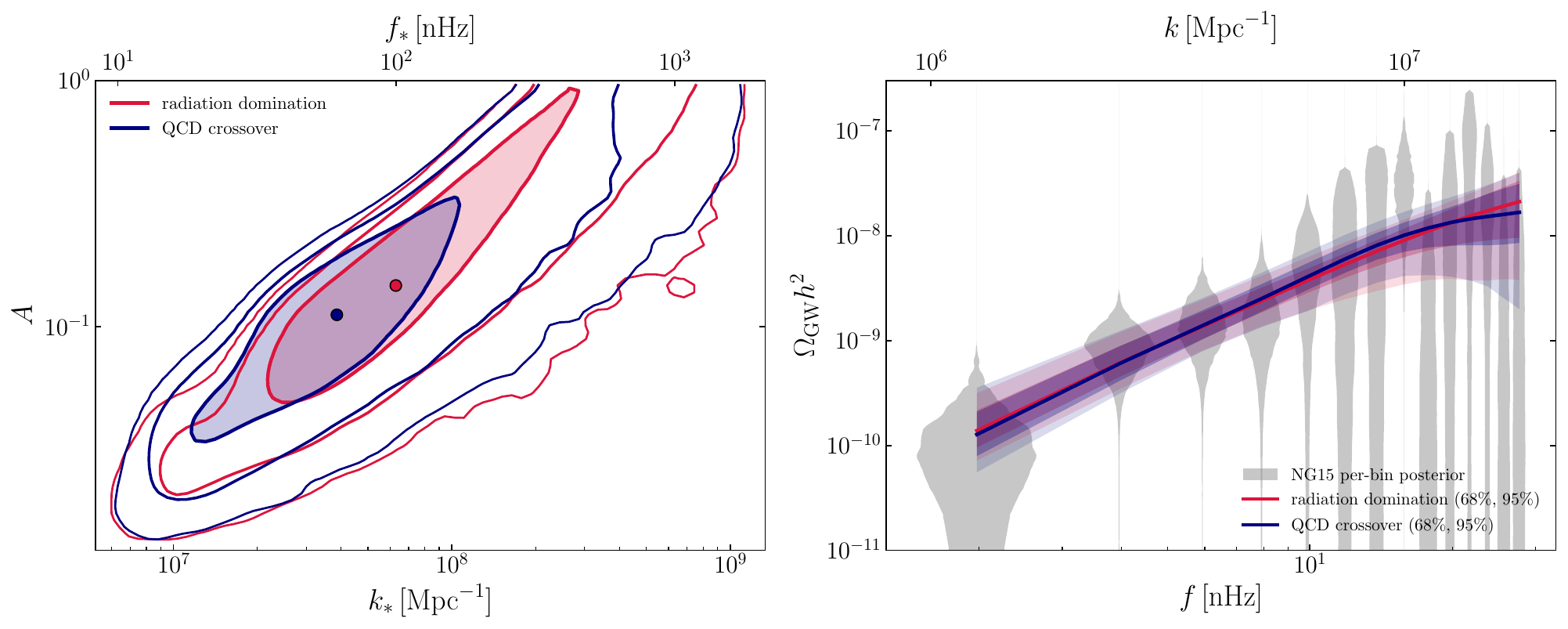}
\caption{\emph{Left}: posterior on $(\log_{10}k_\ast,\log_{10}A)$ for the
broken-power-law ${\cal P}_\zeta$ of Eq.~\eqref{eq:bpl} fitted to the NANOGrav
15-year data, with the exact-RD kernel (red) and the QCD-crossover kernel
(dark blue); dots mark the medians and the contours indicate $1, 2, 3\sigma$  C.L. respectively. 
\emph{Right}: the per-bin NG15 free spectra posteriors
(gray violins, width $\propto$ the bin's posterior) against the
posterior-predictive $\Omega_{\rm GW}h^2$ of each fit, i.e.\
Eq.~\eqref{eq:bpl} propagated through the full RD and QCD chains and
summarized by its $1, 2\sigma$  C.L. bands.}
\label{fig:ptapost}
\end{figure*}

Figure~\ref{fig:ptapost} shows the resulting two-dimensional posterior on
$(\log_{10}k_\ast,\log_{10}A)$. Both fits recover the same strongly
correlated degeneracy --- the PTA band samples only the rising infrared part of a spectrum peaked above it, so a larger $k_\ast$ is compensated by
a larger $A$ --- and both describe the data equally well. Including the
crossover moves the posterior to \emph{lower} amplitude and \emph{lower}
peak scale,
\begin{equation}
\Delta\log_{10}A=-0.09 ,
\qquad
\Delta\log_{10}k_\ast=-0.19 ,
\end{equation}
that is $-0.26\sigma$ and $-0.53\sigma$ in units of the RD $1\sigma$
widths. The direction follows directly from the kernel ratio discussed
above: the QCD kernel is enhanced at the low-frequency end of the NANOGrav
band and suppressed at the high-frequency end, so at fixed ${\cal P}_\zeta$ it
produces a \emph{flatter} background across the band; reproducing the
observed rise then requires slightly less amplitude and a peak slightly
closer to the data. The shift is a fraction of the current statistical
width, so with NANOGrav 15-year data alone the crossover does not change
the qualitative conclusions one would draw about $A$ and $k_\ast$.
This kind of bias would matter once the statistical errors would shrink at least by a factor ${\cal O}(2)$. 
Since it is a \emph{correlated} shift along the degeneracy direction, it is expected to become relevant for derived quantities such as the PBH abundance, which depend
sensitively on the inferred peak location and height. We expect the QCD modulation to become more important as additional pulsars improve sensitivity to a potential SIGW signal in the combined datasets \cite{Cecchini:2025oks} (see also Ref.~\cite{Babak:2024yhu}).

\section{Conclusions}\label{sec:conclusions}

The frequencies probed by current PTA datasets correspond to modes that
re-entered the horizon around the QCD crossover, so any SIGW interpretation of the observed signal is unavoidably sensitive
to the softening of the equation of state at that epoch. We solved the
coupled tensor-scalar system across the SM thermal history
and showed that the crossover leaves an imprint on the induced spectrum
that is not a small correction: across the PTA frequency window the
height of the spectrum changes by up to $\approx55\%$ relative to the
RD expectation, suppressed or enhanced depending on
whether the source modes cross the horizon before or after the transition. 

Fitting
the NANOGrav 15-year data with a broken-power-law curvature spectrum, we
found that including this effect shifts the inferred peak amplitude and
scale a fraction of the current uncertainties along their shared degeneracy. This systematic bias will increasingly become visible  once the statistical uncertainty on these
parameters shrinks further, and may already be relevant
for any PBH abundance derived from the same fit. 

The main deliverables of this work are the
tabulated kernels,
which we make available so that this SM correction can be
folded into future PTA analyses of a putative SIGW signal.
The tabulated kernels, together with the code used to compute them and to
produce all the figures of this work, are publicly available at
\url{https://github.com/gfranciolini/QCD-SIGW}, and can be used to
replace the RD kernel in existing PTA pipelines.
Evaluating the induced spectrum at the $14$ NANOGrav
frequency bins from the tabulated kernel, for one point in the
$(A,k_\ast,\alpha,\beta)$ parameter space of Sec.~\ref{sec:pta}, takes
about $1\,$ms on a single CPU core; the full $32$-walker, $300\,000$-step
production posteriors of Fig.~\ref{fig:ptapost} complete in about
$80\,$minutes.
We comment on the fact that these kernels can also be used in extensions of the SIGW model including primordial non-Gaussianity in
Appendix~\ref{app:ng}.

We leave an updated comparison between the recent PTA dataset and the PBH overproduction bounds for future work. These bounds can severely constrain the interpretation of the signal as originating from SIGWs \cite{Saito:2008jc,Bugaev:2009kq,Bugaev:2010bb,Byrnes:2018txb,Gow:2020bzo,Franciolini:2023pbf,Dandoy:2023jot,Iovino:2024tyg}. Additionally, as the PTA scales are associated with the formation of stellar mass PBHs, these bounds may be complementary to LIGO/Virgo searches, in which the QCD effect was already included (e.g. \cite{Franciolini:2022tfm,Escriva:2022bwe}).

Finally, let us mention that many theories beyond the Standard Model (BSM), such as supersymmetry~\cite{Coleman:1967ad,Fayet:1977yc,Farrar:1978xj} and composite Higgs~\cite{Kaplan:1983fs,Kaplan:1983sm} models, predict a large number of additional degrees of freedom. When such degrees of freedom become non-relativistic, there are expected to be corresponding softenings of the equation of state at higher energy scales, and thus at higher frequencies \cite{Lu:2022yuc,Escriva:2023nzn,Escriva:2024ivo,Pritchard:2025pcn}. This offers future gravitational wave observatories, for example LISA~\cite{Baker:2019nia,LISACosmologyWorkingGroup:2025vdz} and the Einstein Telescope~\cite{Punturo:2010zz}, the prospect of probing such BSM models; the techniques developed in this paper will be used to do this in a future study.

\let\oldaddcontentsline\addcontentsline
\renewcommand{\addcontentsline}[3]{}
\begin{acknowledgments}
We thank Hardi Veerm\"ae for collaboration at the early stages of this work. 
We thank Davide Racco and Fabrizio Rompineve for discussions and comments on an earlier version of the draft. 
We acknowledge the use of ChatGPT and Claude Code to help with writing and running code, exporting figures, and draft proofreading. The authors directed,
checked, and revised all outputs included in this work.
G.F. acknowledges support from the Italian Ministero dell'Università e della Ricerca through the FIS 3 project FIS-2024-02546 “PRIMAVERA” (CUP C53C25001070001), the Departments of Excellence grant 2023–2027 “Quantum Frontier”, as well as from Istituto Nazionale di Fisica Nucleare (INFN) through the Theoretical Astroparticle Physics (TAsP) project. 
X.P. is supported by an STFC studentship and STFC grant \texttt{ST/X001040/1}. X.P. would like to thank Rikkyo University for its hospitality during much of this work. 
Y.T. is supported by JSPS KAKENHI Grant
No.~JP24K07047.
\end{acknowledgments}

\let\addcontentsline\oldaddcontentsline

\appendix
\onecolumngrid



\section{Inclusion of non-Gaussian corrections}
\label{app:ng}

In this appendix, we show that the tabulated kernels can also be used in models which include non-Gaussian corrections to the SIGW signal. 
We summarize here the formulas used to compute the spectrum of SIGWs when non-Gaussianities are included, referring the reader to Refs.~\cite{Unal:2018yaa, Cai:2018dig, Yuan:2020iwf, Adshead:2021hnm, Abe:2022xur, Chang:2022nzu, Garcia-Saenz:2022tzu, Li:2023qua,Perna:2024ehx,Yuan:2023ofl,Li:2023xtl,Chang:2023aba,Wang:2023sij,Zeng:2025cer,Li:2025met,Caravano:2026hca}
for more details.
For presentation purposes, we restrict our discussion to the next-to-leading order corrections in non-Gaussianity to the tensor power spectrum by assuming that the curvature perturbation can be expanded in the form
$\zeta = \zeta_{\rm G} + F_{\rm NL}\zeta_{\rm G}^2$,
where $\zeta_{\rm G}$ is a Gaussian field. In the alternative convention, $F_{\rm NL}=(3/5)f_{\rm NL}$.

The non-Gaussian corrections to the GW power spectrum can then be written as
\begin{align}\label{eq:Omegabar-total}
     \frac{h^2 \Omega_{\rm GW}(k)}{h^2 \Omega_{r,0}}
    =   c_g(k)
\overline{
\left  (
        {\cal P}^\text{\tiny G}_h(k)
    +   {\cal P}^{F^2_\text{\tiny NL}}_h(k)
\right )
}\, + \cdots ,
\end{align}
where the two terms scale as $A_\zeta^2$ and $A_\zeta^3F_{\rm NL}^2$
respectively, the former given in Eq.~\eqref{eq:P_h_ts}. Throughout, we work to $\mathcal{O}(F_{\rm NL}^2)$ and do not evaluate the
${\cal O}(F_{\rm NL}^4)$ terms for simplicity.

Following the discussion in Ref.~\cite{Perna:2024ehx}, the higher-order terms in Eq.~\eqref{eq:Omegabar-total} are the sum of the three non-vanishing $F_{\rm NL}^2$ diagrams (``t'', ``u'' and ``hybrid'' in the notation of Ref.~\cite{Perna:2024ehx}; the fourth, ``s'', vanishes identically by the angular integration). We quote them here converted to our normalization of $\Omega_{\rm GW}$ and of the kernel, related to those of Ref.~\cite{Perna:2024ehx} by $\overline{{\cal P}_h}=24\,\Omega_{\rm GW}/(k/a H)^2$ and $J=4\tilde J$ respectively, to give
\begin{subequations}
\begin{align}
{\cal P}^{F^2_\text{\tiny NL}}_h(k,\eta_{\rm c})
& = \frac{F_{\rm NL}^2}{8}
\prod_{i=1}^2
\biggl[\int_0^\infty \dd{t_i} \int_{-1}^1 \dd{s_i} u_i v_i \biggr]
\Bigg\{
\frac{J^2 (k, t_1,s_1) }{(u_1 v_1 u_2 v_2)^3}
{\cal P}_\zeta  (v_1 v_2 k)
{\cal P}_\zeta  (u_1 k)
{\cal P}_\zeta  (v_1 u_2 k)
\quad\text{(hybrid)}
\nonumber
\\
&+ \frac{1}{\pi}\int_0^{2\pi} \dd{\varphi_{12}}
\cos (2\varphi_{12} )
J (k, t_1,s_1) J (k, t_2,s_2)
\frac{{\cal P}_\zeta (v_2 k)}{v_2^3}
\frac{{\cal P}_\zeta (u_2 k)}{u_2^3}
\frac{{\cal P}_\zeta  (w_{12} k)}{w_{12}^3}
\quad\text{(t)}
\nonumber
\\
&+ \frac{1}{\pi}\int_0^{2\pi} \dd{\varphi_{12}}
\cos (2\varphi_{12} )
J (k, t_1,s_1) J (k, t_2,s_2)
\frac{{\cal P}_\zeta (v_1 k)}{v_1^3}
\frac{{\cal P}_\zeta (v_2 k)}{v_2^3}
\frac{{\cal P}_\zeta  (\bar w_{12} k)}{\bar w_{12}^3}
\quad\text{(u)}
\Bigg\}\, ,
\label{eq:Omega-C}
\end{align}
\end{subequations}
where we defined $s_i=u_i-v_i$, $t_i=u_i+v_i -1 $,
$z_i = \frac{1}{2}\left [1-s_i(t_i+1)\right ]$, and, for a pair $(i,j)$
subtending an azimuthal angle $\varphi_{ij}$ around $\hat k$,
\begin{subequations}
\begin{eqnarray}
    y_{ij}(\varphi_{ij})&=&
    \frac{\cos\varphi_{ij}}{4}
    \sqrt{
    t_i (t_i+2) (s_i^2-1)
    t_j (t_j+2) (s_j^2-1)
    }
    +\frac{1}{4}
    [1-s_i(t_i+1)][1-s_j(t_j+1)]\ ,
    \\
    w_{ij}&=&\sqrt{v_i^2+v_j^2-2y_{ij}}\, ,
    \qquad
    \bar w_{ij}\ =\ \sqrt{1+v_i^2+v_j^2-2z_i-2z_j+2y_{ij}}\ ,
\end{eqnarray}
\end{subequations}
so that $w_{ij}=|\vec q_i - \vec q_j|/k$ and $\bar w_{ij}=|\vec k -\vec q_i-\vec q_j|/k$.
Similar but more complex expressions can be obtained by accounting for the higher-order terms proportional to $g_\text{\tiny NL}$ etc.
Notice, however, that no additional tables are needed for $s<0$: the reflection $s\to-s$
amounts to $u\leftrightarrow v$, under which the source $f$ of
Eq.~\eqref{eq:source_f} is symmetric, so $I_1$, $I_2$ and $J$ are even in
$s$ on any background and we evaluate them at $|s_i|$. Unlike in
Eq.~\eqref{eq:P_h_ts}, however, the remaining factors of
Eq.~\eqref{eq:Omega-C} --- the arguments of ${\cal P}_\zeta$ and, through
$z_i$, the momenta $w_{12}$ and $\bar w_{12}$ --- are not even in each $s_i$
separately, so the halving used for the Gaussian term does not apply and
each $s_i$ integral runs over the full range $[-1,1]$.

\begin{figure} 
\centering
\includegraphics[width=0.8\linewidth]{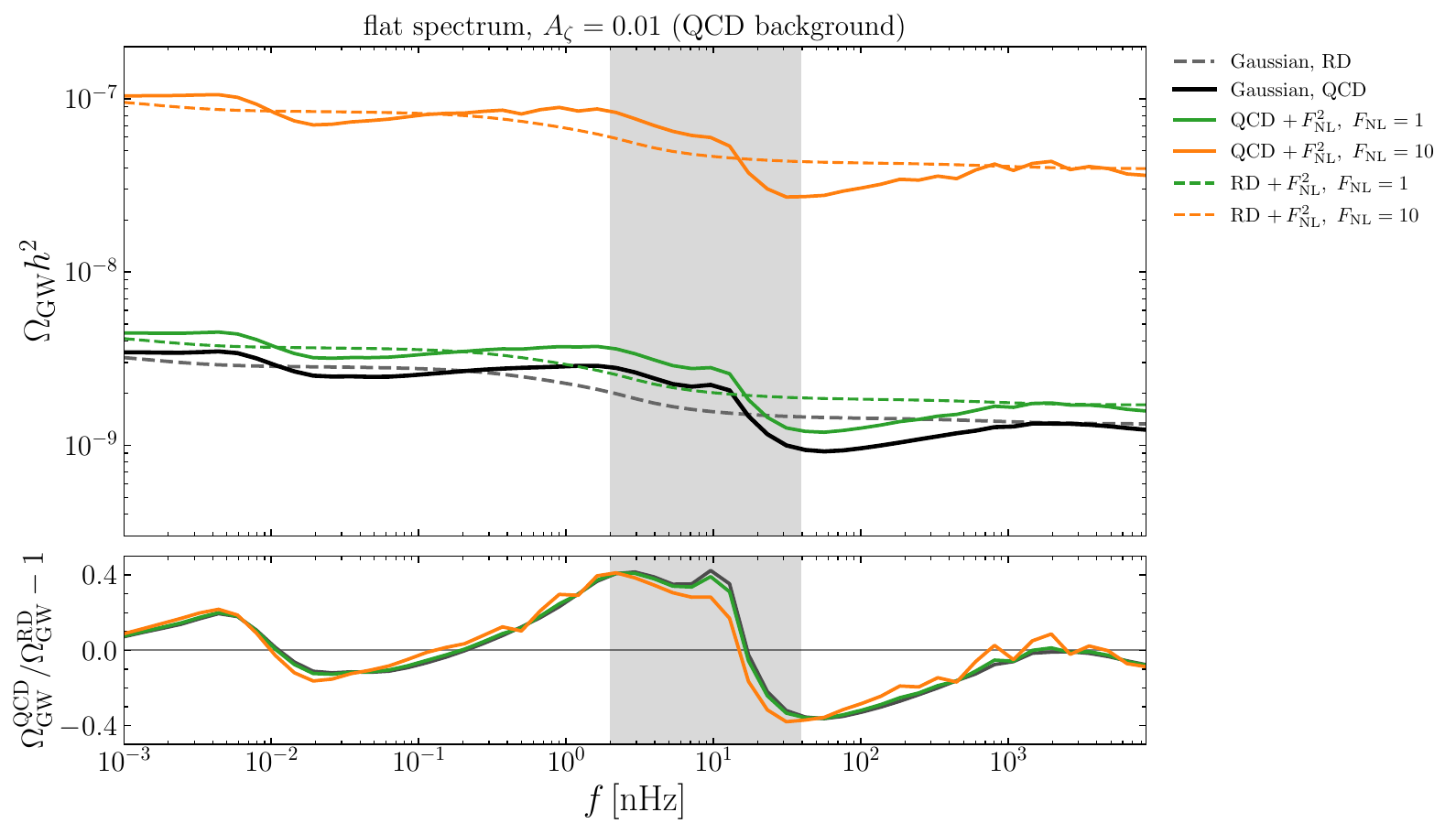}
\caption{\emph{Upper panel}: 
same as Fig.~\ref{fig:flat} assuming $A_\zeta=0.01$.
We show the Gaussian
spectrum on the QCD kernel and on the exact-RD kernel (solid black and
dashed gray, respectively, the latter rescaled by the same $c_g(k)$ as the
QCD curve), each with the $F_{\rm NL}^2$ correction of
Eq.~\eqref{eq:Omega-C} added for $F_{\rm NL}=1$ and $10$ (solid for the QCD
background, dashed of the same color for RD). \emph{Lower panel}: fractional difference between the QCD and RD curves at matched $F_{\rm NL}$ (Gaussian $F_{\rm NL}=0$,
$F_{\rm NL}=1$, $F_{\rm NL}=10$).}
\label{fig:ngflat}
\end{figure}

\begin{figure} 
\centering
\includegraphics[width=\linewidth]{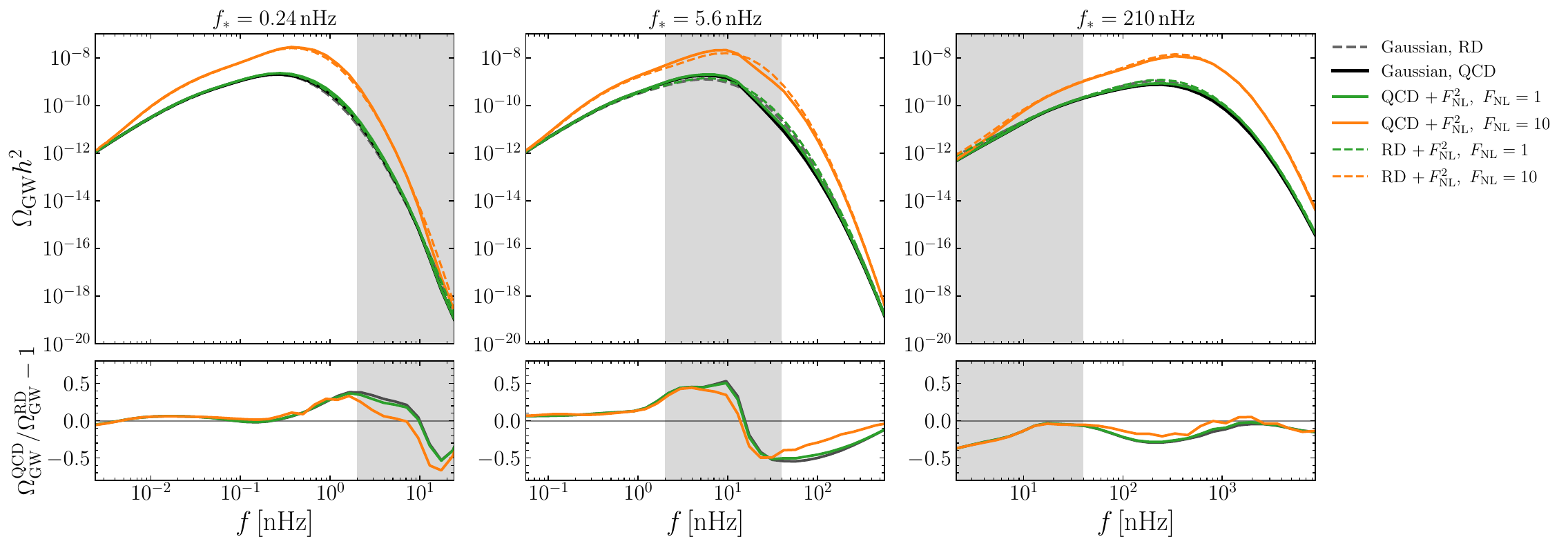}
\caption{\emph{Upper row}: 
same as Fig.~\ref{fig:lognormal_and_peakratio} (lognormal, $\sigma=0.4$, QCD
background) for three peak frequencies $f_\ast$, assuming $A_\zeta=0.01$, as
in Fig.~\ref{fig:ngflat}. \emph{Lower row}: fractional difference between the
QCD and RD curves at matched $F_{\rm NL}$, one panel per $f_\ast$ directly
below its spectrum, common $y$-axis range throughout.}
\label{fig:nglognormal}
\end{figure}

We evaluate Eq.~\eqref{eq:Omega-C} numerically and add it to the Gaussian
spectrum of Sec.~\ref{sec:results}, at fixed $A_\zeta=0.01$ and
$F_{\rm NL}=1$ and $10$, for the two spectral shapes of Figs.~\ref{fig:flat}
and~\ref{fig:lognormal_and_peakratio}. The ``t'' and ``u'' diagrams
require the kernel's cross-oscillation-average at two different $(t,s)$
points, Eq.~\eqref{eq:osc_avg}: we evaluate it exactly on the QCD
background from the tabulated $I_1(k,t,s)$, $I_2(k,t,s)$ and
$\overline{g_{1k}^2}$, $\overline{g_{2k}^2}$, $\overline{g_{1k}g_{2k}}$, and on
the exact-RD background from the closed form of
Ref.~\cite{Perna:2024ehx}. The $(t_1,s_1,t_2,s_2,\varphi_{12})$ integral
is evaluated by deterministic quadrature, clustered around the kernel
resonance.\footnote{We note the presence of small numerical noise in the result. This can be reduced by enhancing the number of points in the tabulated kernels, or by interpolating the kernels and performing the integration on a flexible grid, e.g. using monte-carlo integration techniques. }

For the flat spectrum, two loop-momentum configurations in the bulk of the
integration domain --- $u,v\to0$ in the ``hybrid'' diagram, and the
collinear limit $\vec q_1\to\vec q_2$ (or $\vec k-\vec q_2$) in
``t''/``u'' --- are not integrable once ${\cal P}_\zeta$ fails to suppress
them, as happens for an exactly scale-invariant spectrum; we impose an
explicit infrared floor of $0.1$ on the relevant ratios to obtain a reproducible
number, and the flat-spectrum curves of Fig.~\ref{fig:ngflat} should be
read only for their qualitative trend. The lognormal spectrum of
Fig.~\ref{fig:nglognormal} suppresses both configurations near its peak
and needs no such regulator. This issue is not present in realistic enhanced spectra, necessarily featuring a growth from the large-scale, nearly scale-invariant power spectrum observed through the cosmic microwave background and the large-scale structure. 

Figure~\ref{fig:ngflat} shows the flat-spectrum comparison of
Fig.~\ref{fig:flat}. The upper panel shows the Gaussian spectrum on both backgrounds
together with the $F_{\rm NL}^2$-corrected curves; the lower panel
isolates the QCD-vs-RD effect as a fractional difference at each
$F_{\rm NL}$. 
Figure~\ref{fig:nglognormal} repeats the same comparison for
three of the four peak frequencies of Fig.~\ref{fig:lognormal_and_peakratio}, with a common
$y$-axis range in both rows for direct comparison across panels.

As expected, the spectra featuring a large $F_{\rm NL}$ show a larger overall amplitude (orange in both plots). It is interesting to inspect the fractional deviation induced by QCD. On top of an overall amplitude change, the QCD crossover also produces a marked change in the spectral shape. 
For the
lognormal spectrum, $A_\zeta F_{\rm NL}^2$ ranges from $10^{-2}$
($F_{\rm NL}=1$, safely perturbative) to $1$ ($F_{\rm NL}=10$, marginal);
the correction itself grows from negligible near the peak to an
$\mathcal O(1)$ distortion already at $F_{\rm NL}=10$, most
visibly for peak frequencies below the PTA band.
These results show that the QCD corrections should be included also when accounting for non-Gaussian corrections to SIGW spectra in order to have precise predictions.

\let\oldaddcontentsline\addcontentsline
\renewcommand{\addcontentsline}[3]{}
\bibliography{main}

@article{Li:2023qua,
    author = "Li, Jun-Peng and Wang, Sai and Zhao, Zhi-Chao and Kohri, Kazunori",
    title = "{Primordial non-Gaussianity f $_{NL}$ and anisotropies in scalar-induced gravitational waves}",
    eprint = "2305.19950",
    archivePrefix = "arXiv",
    primaryClass = "astro-ph.CO",
    reportNumber = "KEK-Cosmo-0315, KEK-TH-2531, KEK-QUP-2023-0012",
    doi = "10.1088/1475-7516/2023/10/056",
    journal = "JCAP",
    volume = "10",
    pages = "056",
    year = "2023"
}

@article{Tomita:1975kj,
    author = "Tomita, Kenji",
    title = "{Evolution of Irregularities in a Chaotic Early Universe}",
    reportNumber = "RRK 75-3",
    doi = "10.1143/PTP.54.730",
    journal = "Prog. Theor. Phys.",
    volume = "54",
    pages = "730",
    year = "1975"
}

@article{Matarrese:1993zf,
    author = "Matarrese, Sabino and Pantano, Ornella and Saez, Diego",
    title = "{General relativistic dynamics of irrotational dust: Cosmological implications}",
    eprint = "astro-ph/9310036",
    archivePrefix = "arXiv",
    reportNumber = "DFPD-93-A-67",
    doi = "10.1103/PhysRevLett.72.320",
    journal = "Phys. Rev. Lett.",
    volume = "72",
    pages = "320--323",
    year = "1994"
}

@article{Acquaviva:2002ud,
    author = "Acquaviva, Viviana and Bartolo, Nicola and Matarrese, Sabino and Riotto, Antonio",
    title = "{Second order cosmological perturbations from inflation}",
    eprint = "astro-ph/0209156",
    archivePrefix = "arXiv",
    reportNumber = "DFPD-A-02-21",
    doi = "10.1016/S0550-3213(03)00550-9",
    journal = "Nucl. Phys. B",
    volume = "667",
    pages = "119--148",
    year = "2003"
}

@article{Mollerach:2003nq,
    author = "Mollerach, Silvia and Harari, Diego and Matarrese, Sabino",
    title = "{CMB polarization from secondary vector and tensor modes}",
    eprint = "astro-ph/0310711",
    archivePrefix = "arXiv",
    doi = "10.1103/PhysRevD.69.063002",
    journal = "Phys. Rev. D",
    volume = "69",
    pages = "063002",
    year = "2004"
}

@article{Ananda:2006af,
    author = "Ananda, Kishore N. and Clarkson, Chris and Wands, David",
    title = "{The Cosmological gravitational wave background from primordial density perturbations}",
    eprint = "gr-qc/0612013",
    archivePrefix = "arXiv",
    doi = "10.1103/PhysRevD.75.123518",
    journal = "Phys. Rev. D",
    volume = "75",
    pages = "123518",
    year = "2007"
}

@article{Baumann:2007zm,
    author = "Baumann, Daniel and Steinhardt, Paul J. and Takahashi, Keitaro and Ichiki, Kiyotomo",
    title = "{Gravitational Wave Spectrum Induced by Primordial Scalar Perturbations}",
    eprint = "hep-th/0703290",
    archivePrefix = "arXiv",
    doi = "10.1103/PhysRevD.76.084019",
    journal = "Phys. Rev. D",
    volume = "76",
    pages = "084019",
    year = "2007"
}

@book{Mukhanov:2005sc,
    author = "Mukhanov, V.",
    title = "{Physical Foundations of Cosmology}",
    doi = "10.1017/CBO9780511790553",
    isbn = "978-0-521-56398-7",
    publisher = "Cambridge University Press",
    address = "Oxford",
    year = "2005"
}

@article{Espinosa:2018eve,
    author = "Espinosa, Jos\'e Ram\'on and Racco, Davide and Riotto, Antonio",
    title = "{A Cosmological Signature of the SM Higgs Instability: Gravitational Waves}",
    eprint = "1804.07732",
    archivePrefix = "arXiv",
    primaryClass = "hep-ph",
    doi = "10.1088/1475-7516/2018/09/012",
    journal = "JCAP",
    volume = "09",
    pages = "012",
    year = "2018"
}

@article{Kohri:2018awv,
    author = "Kohri, Kazunori and Terada, Takahiro",
    title = "{Semianalytic calculation of gravitational wave spectrum nonlinearly induced from primordial curvature perturbations}",
    eprint = "1804.08577",
    archivePrefix = "arXiv",
    primaryClass = "gr-qc",
    reportNumber = "KEK-TH-2046, KEK-COSMO-223",
    doi = "10.1103/PhysRevD.97.123532",
    journal = "Phys. Rev. D",
    volume = "97",
    number = "12",
    pages = "123532",
    year = "2018"
}

@article{Maggiore:1999vm,
    author = "Maggiore, Michele",
    title = "{Gravitational wave experiments and early universe cosmology}",
    eprint = "gr-qc/9909001",
    archivePrefix = "arXiv",
    reportNumber = "IFUP-TH-20-99",
    doi = "10.1016/S0370-1573(99)00102-7",
    journal = "Phys. Rept.",
    volume = "331",
    pages = "283--367",
    year = "2000"
}

@article{Abe:2020sqb,
    author = "Abe, Katsuya T. and Tada, Yuichiro and Ueda, Ikumi",
    title = "{Induced gravitational waves as a cosmological probe of the sound speed during the QCD phase transition}",
    eprint = "2010.06193",
    archivePrefix = "arXiv",
    primaryClass = "astro-ph.CO",
    doi = "10.1088/1475-7516/2021/06/048",
    journal = "JCAP",
    volume = "06",
    pages = "048",
    year = "2021"
}

@article{Unal:2018yaa,
    author = "Unal, Caner",
    title = "{Imprints of Primordial Non-Gaussianity on Gravitational Wave Spectrum}",
    eprint = "1811.09151",
    archivePrefix = "arXiv",
    primaryClass = "astro-ph.CO",
    doi = "10.1103/PhysRevD.99.041301",
    journal = "Phys. Rev. D",
    volume = "99",
    number = "4",
    pages = "041301",
    year = "2019"
}

@article{Cai:2018dig,
    author = "Cai, Rong-gen and Pi, Shi and Sasaki, Misao",
    title = "{Gravitational Waves Induced by non-Gaussian Scalar Perturbations}",
    eprint = "1810.11000",
    archivePrefix = "arXiv",
    primaryClass = "astro-ph.CO",
    reportNumber = "IPMU18-0172, YITP-18-114",
    doi = "10.1103/PhysRevLett.122.201101",
    journal = "Phys. Rev. Lett.",
    volume = "122",
    number = "20",
    pages = "201101",
    year = "2019"
}

@article{Yuan:2020iwf,
    author = "Yuan, Chen and Huang, Qing-Guo",
    title = "{Gravitational waves induced by the local-type non-Gaussian curvature perturbations}",
    eprint = "2007.10686",
    archivePrefix = "arXiv",
    primaryClass = "astro-ph.CO",
    doi = "10.1016/j.physletb.2021.136606",
    journal = "Phys. Lett. B",
    volume = "821",
    pages = "136606",
    year = "2021"
}

@article{Adshead:2021hnm,
    author = "Adshead, Peter and Lozanov, Kaloian D. and Weiner, Zachary J.",
    title = "{Non-Gaussianity and the induced gravitational wave background}",
    eprint = "2105.01659",
    archivePrefix = "arXiv",
    primaryClass = "astro-ph.CO",
    doi = "10.1088/1475-7516/2021/10/080",
    journal = "JCAP",
    volume = "10",
    pages = "080",
    year = "2021"
}

@article{Abe:2022xur,
    author = "Abe, Katsuya T. and Inui, Ryoto and Tada, Yuichiro and Yokoyama, Shuichiro",
    title = "{Primordial black holes and gravitational waves induced by exponential-tailed perturbations}",
    eprint = "2209.13891",
    archivePrefix = "arXiv",
    primaryClass = "astro-ph.CO",
    doi = "10.1088/1475-7516/2023/05/044",
    journal = "JCAP",
    volume = "05",
    pages = "044",
    year = "2023"
}

@article{Chang:2022nzu,
    author = "Chang, Zhe and Kuang, Yu-Ting and Zhang, Xukun and Zhou, Jing-Zhi",
    title = "{Primordial black holes and third order scalar induced gravitational waves*}",
    eprint = "2209.12404",
    archivePrefix = "arXiv",
    primaryClass = "astro-ph.CO",
    doi = "10.1088/1674-1137/acc649",
    journal = "Chin. Phys. C",
    volume = "47",
    number = "5",
    pages = "055104",
    year = "2023"
}

@article{Garcia-Saenz:2022tzu,
    author = "Garcia-Saenz, Sebastian and Pinol, Lucas and Renaux-Petel, S\'ebastien and Werth, Denis",
    title = "{No-go theorem for scalar-trispectrum-induced gravitational waves}",
    eprint = "2207.14267",
    archivePrefix = "arXiv",
    primaryClass = "astro-ph.CO",
    doi = "10.1088/1475-7516/2023/03/057",
    journal = "JCAP",
    volume = "03",
    pages = "057",
    year = "2023"
}

@article{Yuan:2023ofl,
    author = "Yuan, Chen and Meng, De-Shuang and Huang, Qing-Guo",
    title = "{Full analysis of the scalar-induced gravitational waves for the curvature perturbation with local-type non-Gaussianities}",
    eprint = "2308.07155",
    archivePrefix = "arXiv",
    primaryClass = "astro-ph.CO",
    doi = "10.1088/1475-7516/2023/12/036",
    journal = "JCAP",
    volume = "12",
    pages = "036",
    year = "2023"
}

@article{Yuan:2019wwo,
    author = "Yuan, Chen and Chen, Zu-Cheng and Huang, Qing-Guo",
    title = "{Log-dependent slope of scalar induced gravitational waves in the infrared regions}",
    eprint = "1910.09099",
    archivePrefix = "arXiv",
    primaryClass = "astro-ph.CO",
    doi = "10.1103/PhysRevD.101.043019",
    journal = "Phys. Rev. D",
    volume = "101",
    number = "4",
    pages = "4",
    year = "2020"
}

@article{Cai:2019cdl,
    author = "Cai, Rong-Gen and Pi, Shi and Sasaki, Misao",
    title = "{Universal infrared scaling of gravitational wave background spectra}",
    eprint = "1909.13728",
    archivePrefix = "arXiv",
    primaryClass = "astro-ph.CO",
    reportNumber = "IPMU19-0135, YITP-19-88",
    doi = "10.1103/PhysRevD.102.083528",
    journal = "Phys. Rev. D",
    volume = "102",
    number = "8",
    pages = "083528",
    year = "2020"
}

@article{Tomita:1967wkp,
    author = "Tomita, Kenji",
    title = "{Non-Linear Theory of Gravitational Instability in the Expanding Universe}",
    doi = "10.1143/PTP.37.831",
    journal = "Prog. Theor. Phys.",
    volume = "37",
    number = "5",
    pages = "831--846",
    year = "1967"
}

@article{Perna:2024ehx,
    author = "Perna, Gabriele and Testini, Chiara and Ricciardone, Angelo and Matarrese, Sabino",
    title = "{Fully non-Gaussian Scalar-Induced Gravitational Waves}",
    eprint = "2403.06962",
    archivePrefix = "arXiv",
    primaryClass = "astro-ph.CO",
    doi = "10.1088/1475-7516/2024/05/086",
    journal = "JCAP",
    volume = "05",
    pages = "086",
    year = "2024"
}

@article{Li:2023xtl,
    author = "Li, Jun-Peng and Wang, Sai and Zhao, Zhi-Chao and Kohri, Kazunori",
    title = "{Complete analysis of the background and anisotropies of scalar-induced gravitational waves: primordial non-Gaussianity f $_{NL}$ and g $_{NL}$ considered}",
    eprint = "2309.07792",
    archivePrefix = "arXiv",
    primaryClass = "astro-ph.CO",
    reportNumber = "KEK-Cosmo-0326, KEK-TH-2556, KEK-QUP-2023-0024, KEK-QUP-2023-0024,
  https://github.com/Zhi-ChaoZhao/sigw{\_}class",
    doi = "10.1088/1475-7516/2024/06/039",
    journal = "JCAP",
    volume = "06",
    pages = "039",
    year = "2024"
}

@article{LISACosmologyWorkingGroup:2024hsc,
    author = "Braglia, Matteo and others",
    collaboration = "LISA Cosmology Working Group",
    title = "{Gravitational waves from inflation in LISA: reconstruction pipeline and physics interpretation}",
    eprint = "2407.04356",
    archivePrefix = "arXiv",
    primaryClass = "astro-ph.CO",
    reportNumber = "LISA-COSWG-24-03, CERN-TH-2024-072",
    doi = "10.1088/1475-7516/2024/11/032",
    journal = "JCAP",
    volume = "11",
    pages = "032",
    year = "2024"
}

@article{Madge:2023cak,
    author = "Madge, Eric and Morgante, Enrico and Puchades-Ib{\'a}{\~n}ez, Cristina and Ramberg, Nicklas and Ratzinger, Wolfram and Schenk, Sebastian and Schwaller, Pedro",
    title = "{Primordial gravitational waves in the nano-Hertz regime and PTA data {\textemdash} towards solving the GW inverse problem}",
    eprint = "2306.14856",
    archivePrefix = "arXiv",
    primaryClass = "hep-ph",
    reportNumber = "MITP-23-029",
    doi = "10.1007/JHEP10(2023)171",
    journal = "JHEP",
    volume = "10",
    pages = "171",
    year = "2023"
}

@article{EPTA:2023xxk,
    author = "Antoniadis, J. and others",
    collaboration = "EPTA, InPTA",
    title = "{The second data release from the European Pulsar Timing Array - IV. Implications for massive black holes, dark matter, and the early Universe}",
    eprint = "2306.16227",
    archivePrefix = "arXiv",
    primaryClass = "astro-ph.CO",
    doi = "10.1051/0004-6361/202347433",
    journal = "Astron. Astrophys.",
    volume = "685",
    pages = "A94",
    year = "2024"
}

@article{Figueroa:2023zhu,
    author = "Figueroa, Daniel G. and Pieroni, Mauro and Ricciardone, Angelo and Simakachorn, Peera",
    title = "{Cosmological Background Interpretation of Pulsar Timing Array Data}",
    eprint = "2307.02399",
    archivePrefix = "arXiv",
    primaryClass = "astro-ph.CO",
    reportNumber = "CERN-TH-2023-132",
    doi = "10.1103/PhysRevLett.132.171002",
    journal = "Phys. Rev. Lett.",
    volume = "132",
    number = "17",
    pages = "171002",
    year = "2024"
}

@article{Ellis:2023oxs,
    author = {Ellis, John and Fairbairn, Malcolm and Franciolini, Gabriele and H{\"u}tsi, Gert and Iovino, Antonio and Lewicki, Marek and Raidal, Martti and Urrutia, Juan and Vaskonen, Ville and Veerm{\"a}e, Hardi},
    title = "{What is the source of the PTA GW signal?}",
    eprint = "2308.08546",
    archivePrefix = "arXiv",
    primaryClass = "astro-ph.CO",
    reportNumber = "KCL-PH-TH/2023-43, CERN-TH-2023-153, AION-REPORT/2023-08",
    doi = "10.1103/PhysRevD.109.023522",
    journal = "Phys. Rev. D",
    volume = "109",
    number = "2",
    pages = "023522",
    year = "2024"
}

@article{Caprini:2024lxj,
    author = "Caprini, Chiara",
    title = "{Strong evidence for the discovery of a gravitational wave background}",
    doi = "10.1038/s42254-024-00711-6",
    journal = "Nature Rev. Phys.",
    volume = "6",
    number = "5",
    pages = "291--293",
    year = "2024"
}

@article{Moore:2021ibq,
    author = "Moore, Christopher J. and Vecchio, Alberto",
    title = "{Ultra-low-frequency gravitational waves from cosmological and astrophysical processes}",
    eprint = "2104.15130",
    archivePrefix = "arXiv",
    primaryClass = "astro-ph.CO",
    doi = "10.1038/s41550-021-01489-8",
    journal = "Nature Astron.",
    volume = "5",
    number = "12",
    pages = "1268--1274",
    year = "2021"
}

@article{Saito:2008jc,
    author = "Saito, Ryo and Yokoyama, Jun'ichi",
    title = "{Gravitational wave background as a probe of the primordial black hole abundance}",
    eprint = "0812.4339",
    archivePrefix = "arXiv",
    primaryClass = "astro-ph",
    reportNumber = "RESCEU-63-08",
    doi = "10.1103/PhysRevLett.102.161101",
    journal = "Phys. Rev. Lett.",
    volume = "102",
    pages = "161101",
    year = "2009",
    note = "[Erratum: Phys.Rev.Lett. 107, 069901 (2011)]"
}

@article{Bugaev:2009kq,
    author = "Bugaev, E. V. and Klimai, P. A.",
    title = "{Bound on induced gravitational wave background from primordial black holes}",
    eprint = "0911.0611",
    archivePrefix = "arXiv",
    primaryClass = "astro-ph.CO",
    doi = "10.1134/S0021364010010017",
    journal = "JETP Lett.",
    volume = "91",
    pages = "1--5",
    year = "2010"
}

@article{Bugaev:2010bb,
    author = "Bugaev, Edgar and Klimai, Peter",
    title = "{Constraints on the induced gravitational wave background from primordial black holes}",
    eprint = "1012.4697",
    archivePrefix = "arXiv",
    primaryClass = "astro-ph.CO",
    doi = "10.1103/PhysRevD.83.083521",
    journal = "Phys. Rev. D",
    volume = "83",
    pages = "083521",
    year = "2011"
}

@article{Byrnes:2018txb,
    author = "Byrnes, Christian T. and Cole, Philippa S. and Patil, Subodh P.",
    title = "{Steepest growth of the power spectrum and primordial black holes}",
    eprint = "1811.11158",
    archivePrefix = "arXiv",
    primaryClass = "astro-ph.CO",
    doi = "10.1088/1475-7516/2019/06/028",
    journal = "JCAP",
    volume = "06",
    pages = "028",
    year = "2019"
}

@article{Gow:2020bzo,
    author = "Gow, Andrew D. and Byrnes, Christian T. and Cole, Philippa S. and Young, Sam",
    title = "{The power spectrum on small scales: Robust constraints and comparing PBH methodologies}",
    eprint = "2008.03289",
    archivePrefix = "arXiv",
    primaryClass = "astro-ph.CO",
    doi = "10.1088/1475-7516/2021/02/002",
    journal = "JCAP",
    volume = "02",
    pages = "002",
    year = "2021"
}

@article{Franciolini:2023pbf,
    author = "Franciolini, Gabriele and Iovino, Junior., Antonio and Vaskonen, Ville and Veermae, Hardi",
    title = "{Recent Gravitational Wave Observation by Pulsar Timing Arrays and Primordial Black Holes: The Importance of Non-Gaussianities}",
    eprint = "2306.17149",
    archivePrefix = "arXiv",
    primaryClass = "astro-ph.CO",
    doi = "10.1103/PhysRevLett.131.201401",
    journal = "Phys. Rev. Lett.",
    volume = "131",
    number = "20",
    pages = "201401",
    year = "2023"
}

@article{Dandoy:2023jot,
    author = "Dandoy, Virgile and Domcke, Valerie and Rompineve, Fabrizio",
    title = "{Search for scalar induced gravitational waves in the international pulsar timing array data release 2 and NANOgrav 12.5 years datasets}",
    eprint = "2302.07901",
    archivePrefix = "arXiv",
    primaryClass = "astro-ph.CO",
    reportNumber = "CERN-TH-2023-027",
    doi = "10.21468/SciPostPhysCore.6.3.060",
    journal = "SciPost Phys. Core",
    volume = "6",
    pages = "060",
    year = "2023"
}

@article{Iovino:2024tyg,
    author = {Iovino, A. J. and Perna, G. and Riotto, A. and Veerm{\"a}e, H.},
    title = "{Curbing PBHs with PTAs}",
    eprint = "2406.20089",
    archivePrefix = "arXiv",
    primaryClass = "astro-ph.CO",
    doi = "10.1088/1475-7516/2024/10/050",
    journal = "JCAP",
    volume = "10",
    pages = "050",
    year = "2024"
}

@article{Zeldovich:1967lct,
    author = "Zel'dovich, Ya. B. and Novikov, I. D.",
    title = "{The Hypothesis of Cores Retarded during Expansion and the Hot Cosmological Model}",
    journal = "Sov. Astron.",
    volume = "10",
    pages = "602",
    year = "1967"
}

@article{Hawking:1971ei,
    author = "Hawking, Stephen",
    title = "{Gravitationally collapsed objects of very low mass}",
    doi = "10.1093/mnras/152.1.75",
    journal = "Mon. Not. Roy. Astron. Soc.",
    volume = "152",
    pages = "75",
    year = "1971"
}

@article{Carr:1974nx,
    author = "Carr, Bernard J. and Hawking, S. W.",
    title = "{Black holes in the early Universe}",
    doi = "10.1093/mnras/168.2.399",
    journal = "Mon. Not. Roy. Astron. Soc.",
    volume = "168",
    pages = "399--415",
    year = "1974"
}

@article{Carr:1975qj,
    author = "Carr, Bernard J.",
    title = "{The Primordial black hole mass spectrum}",
    doi = "10.1086/153853",
    journal = "Astrophys. J.",
    volume = "201",
    pages = "1--19",
    year = "1975"
}

@article{LISACosmologyWorkingGroup:2025vdz,
    author = "Gammal, Jonas El and others",
    collaboration = "LISA Cosmology Working Group",
    title = "{Reconstructing primordial curvature perturbations via scalar-induced gravitational waves with LISA}",
    eprint = "2501.11320",
    archivePrefix = "arXiv",
    primaryClass = "astro-ph.CO",
    reportNumber = "CERN-TH-2024-217",
    doi = "10.1088/1475-7516/2025/05/062",
    journal = "JCAP",
    volume = "05",
    pages = "062",
    year = "2025"
}

@article{Chapline:1975ojl,
    author = "Chapline, George F.",
    title = "{Cosmological effects of primordial black holes}",
    doi = "10.1038/253251a0",
    journal = "Nature",
    volume = "253",
    number = "5489",
    pages = "251--252",
    year = "1975"
}

@book{Byrnes:2025tji,
    editor = "Byrnes, Christian and Franciolini, Gabriele and Harada, Tomohiro and Pani, Paolo and Sasaki, Misao",
    title = "{Primordial Black Holes}",
    doi = "10.1007/978-981-97-8887-3",
    isbn = "978-981--978886-6, 978-981--978889-7, 978-981--978887-3",
    publisher = "Springer",
    series = "Springer Series in Astrophysics and Cosmology",
    year = "2025"
}

@article{Babak:2024yhu,
    author = "Babak, Stanislav and Falxa, Mikel and Franciolini, Gabriele and Pieroni, Mauro",
    title = "{Forecasting the sensitivity of pulsar timing arrays to gravitational wave backgrounds}",
    eprint = "2404.02864",
    archivePrefix = "arXiv",
    primaryClass = "astro-ph.CO",
    reportNumber = "CERN-TH-2024-039",
    doi = "10.1103/PhysRevD.110.063022",
    journal = "Phys. Rev. D",
    volume = "110",
    number = "6",
    pages = "063022",
    year = "2024"
}

@article{Cecchini:2025oks,
    author = "Cecchini, Chiara and Franciolini, Gabriele and Pieroni, Mauro",
    title = "{Forecasting constraints on scalar-induced gravitational waves with future pulsar timing array observations}",
    eprint = "2503.10805",
    archivePrefix = "arXiv",
    primaryClass = "astro-ph.CO",
    reportNumber = "CERN-TH-2025-045",
    doi = "10.1103/nxx5-gx7d",
    journal = "Phys. Rev. D",
    volume = "111",
    number = "12",
    pages = "123536",
    year = "2025"
}

@article{Caravano:2026hca,
    author = "Caravano, Angelo and Franciolini, Gabriele and Renaux-Petel, S{\'e}bastien",
    title = "{Lattice simulations of scalar-induced gravitational waves from inflation}",
    eprint = "2604.03628",
    archivePrefix = "arXiv",
    primaryClass = "astro-ph.CO",
    doi = "10.1103/rzyb-52jx",
    journal = "Phys. Rev. D",
    volume = "114",
    number = "6",
    pages = "063530",
    year = "2026"
}

@unpublished{Yuan:2026krm,
    author = "Yuan, Chen",
    title = "{Infrared Universality of Scalar Induced Gravitational Waves Beyond Second Order}",
    eprint = "2609.17928",
    archivePrefix = "arXiv",
    primaryClass = "astro-ph.CO",
    month = "9",
    year = "2026"
}

@article{Miles:2024seg,
    author = "Miles, Matthew T. and others",
    title = "{The MeerKAT Pulsar Timing Array: the first search for gravitational waves with the MeerKAT radio telescope}",
    eprint = "2412.01153",
    archivePrefix = "arXiv",
    primaryClass = "astro-ph.HE",
    doi = "10.1093/mnras/stae2571",
    journal = "Mon. Not. Roy. Astron. Soc.",
    volume = "536",
    number = "2",
    pages = "1489--1500",
    year = "2024"
}

@article{Franciolini:2023wjm,
    author = "Franciolini, Gabriele and Racco, Davide and Rompineve, Fabrizio",
    title = "{Footprints of the QCD Crossover on Cosmological Gravitational Waves at Pulsar Timing Arrays}",
    eprint = "2306.17136",
    archivePrefix = "arXiv",
    primaryClass = "astro-ph.CO",
    reportNumber = "CERN-TH-2023-080",
    doi = "10.1103/PhysRevLett.132.081001",
    journal = "Phys. Rev. Lett.",
    volume = "132",
    number = "8",
    pages = "081001",
    year = "2024",
    note = "[Erratum: Phys.Rev.Lett. 133, 189901 (2024)]"
}

@unpublished{Feng:2026xwp,
    author = "Feng, Wan-Zhe and Li, Ao and Zhou, Jing-Zhi",
    title = "{Scalar induced gravitational waves as probes of dark QCD}",
    eprint = "2608.12706",
    archivePrefix = "arXiv",
    primaryClass = "hep-ph",
    month = "8",
    year = "2026"
}

@article{Chang:2023aba,
    author = "Chang, Zhe and Kuang, Yu-Ting and Wu, Di and Zhou, Jing-Zhi and Zhu, Qing-Hua",
    title = "{New constraints on primordial non-Gaussianity from missing two-loop contributions of scalar induced gravitational waves}",
    eprint = "2311.05102",
    archivePrefix = "arXiv",
    primaryClass = "astro-ph.CO",
    doi = "10.1103/PhysRevD.109.L041303",
    journal = "Phys. Rev. D",
    volume = "109",
    number = "4",
    pages = "L041303",
    year = "2024"
}

@article{Wang:2023sij,
    author = "Wang, Sai and Zhao, Zhi-Chao and Zhu, Qing-Hua",
    title = "{Constraints on scalar-induced gravitational waves up to third order from a joint analysis of BBN, CMB, and PTA data}",
    eprint = "2307.03095",
    archivePrefix = "arXiv",
    primaryClass = "astro-ph.CO",
    doi = "10.1103/PhysRevResearch.6.013207",
    journal = "Phys. Rev. Res.",
    volume = "6",
    number = "1",
    pages = "013207",
    year = "2024"
}

@article{NANOGrav,
doi = {10.3847/2041-8213/acdac6},
url = {https://dx.doi.org/10.3847/2041-8213/acdac6},
year = {2023},
month = {jun},
publisher = {The American Astronomical Society},
volume = {951},
number = {1},
pages = {L8},
author = {Gabriella Agazie et al.},
title = {The NANOGrav 15 yr Data Set: Evidence for a Gravitational-wave Background},
journal = {The Astrophysical Journal Letters},
}

@article{NANOGrav1,
doi = {10.3847/2041-8213/acda9a},
url = {https://dx.doi.org/10.3847/2041-8213/acda9a},
year = {2023},
month = {jun},
publisher = {The American Astronomical Society},
volume = {951},
number = {1},
pages = {L9},
author = {Gabriella Agazie et al.},
title = {The NANOGrav 15 yr Data Set: Observations and Timing of 68 Millisecond Pulsars},
journal = {The Astrophysical Journal Letters},
}

@article{EPTA,
   title={The second data release from the European Pulsar Timing Array: III. Search for gravitational wave signals},
   volume={678},
   ISSN={1432-0746},
   url={http://dx.doi.org/10.1051/0004-6361/202346844},
   DOI={10.1051/0004-6361/202346844},
   journal={Astronomy \& Astrophysics},
   publisher={EDP Sciences},
   author={Antoniadis, J. et al.},
   year={2023},
   month=oct, pages={A50} }

@article{EPTA1,
   title={The second data release from the European Pulsar Timing Array: I. The dataset and timing analysis},
   volume={678},
   ISSN={1432-0746},
   url={http://dx.doi.org/10.1051/0004-6361/202346841},
   DOI={10.1051/0004-6361/202346841},
   journal={Astronomy \& Astrophysics},
   publisher={EDP Sciences},
   author={Antoniadis, J. et al.},
   year={2023},
   month=oct, pages={A48} }

@misc{EPTA2,
      title={The second data release from the European Pulsar Timing Array: V. Implications for massive black holes, dark matter and the early Universe}, 
      author={J. Antoniadis et al.},
      year={2023},
      eprint={2306.16227},
      archivePrefix={arXiv},
      primaryClass={astro-ph.CO}}

@article{PPTA,
   title={Search for an Isotropic Gravitational-wave Background with the Parkes Pulsar Timing Array},
   volume={951},
   ISSN={2041-8213},
   url={http://dx.doi.org/10.3847/2041-8213/acdd02},
   DOI={10.3847/2041-8213/acdd02},
   number={1},
   journal={The Astrophysical Journal Letters},
   publisher={American Astronomical Society},
   author={Reardon, Daniel J. et al.},
   year={2023},
   month=jun, pages={L6} }

@misc{PPTA1,
      title={The Parkes Pulsar Timing Array Third Data Release}, 
      author={Andrew Zic et al.},
      year={2023},
      eprint={2306.16230},
      archivePrefix={arXiv},
      primaryClass={astro-ph.HE}
}

@article{PPTA2,
   title={The Gravitational-wave Background Null Hypothesis: Characterizing Noise in Millisecond Pulsar Arrival Times with the Parkes Pulsar Timing Array},
   volume={951},
   ISSN={2041-8213},
   url={http://dx.doi.org/10.3847/2041-8213/acdd03},
   DOI={10.3847/2041-8213/acdd03},
   number={1},
   journal={The Astrophysical Journal Letters},
   publisher={American Astronomical Society},
   author={Reardon, Daniel J. et al.},
   year={2023},
   month=jun, pages={L7} }

@article{CPTA,
   title={Searching for the Nano-Hertz Stochastic Gravitational Wave Background with the Chinese Pulsar Timing Array Data Release I},
   volume={23},
   ISSN={1674-4527},
   url={http://dx.doi.org/10.1088/1674-4527/acdfa5},
   DOI={10.1088/1674-4527/acdfa5},
   number={7},
   journal={Research in Astronomy and Astrophysics},
   publisher={IOP Publishing},
   author={Xu, Heng et al.},
   year={2023},
   month=jun, pages={075024} 
}

@article{NANOGrav:2023hvm,
    author = "Afzal, Adeela and others",
    collaboration = "NANOGrav",
    title = "{The NANOGrav 15 yr Data Set: Search for Signals from New Physics}",
    eprint = "2306.16219",
    archivePrefix = "arXiv",
    primaryClass = "astro-ph.HE",
    reportNumber = "FERMILAB-PUB-23-589-T",
    doi = "10.3847/2041-8213/acdc91",
    journal = "Astrophys. J. Lett.",
    volume = "951",
    number = "1",
    pages = "L11",
    year = "2023",
    note = "[Erratum: Astrophys.J.Lett. 971, L27 (2024), Erratum: Astrophys.J. 971, L27 (2024)]"
}

@article{Saikawa:2018rcs,
    author = "Saikawa, Ken'ichi and Shirai, Satoshi",
    title = "{Primordial gravitational waves, precisely: The role of thermodynamics in the Standard Model}",
    eprint = "1803.01038",
    archivePrefix = "arXiv",
    primaryClass = "hep-ph",
    reportNumber = "IPMU18-0037, MPP-2018-19",
    doi = "10.1088/1475-7516/2018/05/035",
    journal = "JCAP",
    volume = "05",
    pages = "035",
    year = "2018"
}

@article{Englert:1964et,
    author = "Englert, F. and Brout, R.",
    editor = "Taylor, J. C.",
    title = "{Broken Symmetry and the Mass of Gauge Vector Mesons}",
    doi = "10.1103/PhysRevLett.13.321",
    journal = "Phys. Rev. Lett.",
    volume = "13",
    pages = "321--323",
    year = "1964"
}

@article{Higgs:1964pj,
    author = "Higgs, Peter W.",
    editor = "Taylor, J. C.",
    title = "{Broken Symmetries and the Masses of Gauge Bosons}",
    doi = "10.1103/PhysRevLett.13.508",
    journal = "Phys. Rev. Lett.",
    volume = "13",
    pages = "508--509",
    year = "1964"
}

@article{Weinberg:1967tq,
    author = "Weinberg, Steven",
    title = "{A Model of Leptons}",
    doi = "10.1103/PhysRevLett.19.1264",
    journal = "Phys. Rev. Lett.",
    volume = "19",
    pages = "1264--1266",
    year = "1967"
}

@article{Laine:2015kra,
    author = "Laine, M. and Meyer, M.",
    title = "{Standard Model thermodynamics across the electroweak crossover}",
    eprint = "1503.04935",
    archivePrefix = "arXiv",
    primaryClass = "hep-ph",
    doi = "10.1088/1475-7516/2015/07/035",
    journal = "JCAP",
    volume = "07",
    pages = "035",
    year = "2015"
}

@misc{borsanyi1,
      title={Lattice QCD for Cosmology}, 
      author={Sz. Borsanyi and Z. Fodor and K. H. Kampert and S. D. Katz and T. Kawanai and T. G. Kovacs and S. W. Mages and A. Pasztor and F. Pittler and J. Redondo and A. Ringwald and K. K. Szabo},
      year={2016},
      eprint={1606.07494},
      archivePrefix={arXiv},
      primaryClass={hep-lat}
}

@article{PhysRevD.110.030001,
  title = {Review of Particle Physics},
  author = {Navas, S. et al.},
  collaboration = {Particle Data Group Collaboration},
  journal = {Phys. Rev. D},
  volume = {110},
  issue = {3},
  pages = {030001},
  numpages = {5},
  year = {2024},
  month = {Aug},
  publisher = {American Physical Society},
  doi = {10.1103/PhysRevD.110.030001},
  url = {https://link.aps.org/doi/10.1103/PhysRevD.110.030001}
}

@unpublished{Gonin:2026tyu,
    author = {Gonin, Ma{\"e}l and Froustey, Julien and Escriv{\`a}, Albert and Magaraggia, Alberto and Hasinger, G{\"u}nther and K{\"u}hnel, Florian},
    title = "{Primordial Asymmetries, Primordial Equation of State {\&} Primordial Black Holes}",
    eprint = "2608.22240",
    archivePrefix = "arXiv",
    primaryClass = "astro-ph.CO",
    month = "8",
    year = "2026"
}

@article{Lu:2022yuc,
    author = "Lu, Philip and Takhistov, Volodymyr and Fuller, George M.",
    title = "{Signatures of a High Temperature QCD Transition in the Early Universe}",
    eprint = "2212.00156",
    archivePrefix = "arXiv",
    primaryClass = "astro-ph.CO",
    reportNumber = "IPMU22-0064, KEK-QUP-2022-0017, KEK-TH-2476, KEK-Cosmo-0303",
    doi = "10.1103/PhysRevLett.130.221002",
    journal = "Phys. Rev. Lett.",
    volume = "130",
    number = "22",
    pages = "221002",
    year = "2023"
}

@article{Escriva:2023nzn,
    author = "Escriva, Albert and Tada, Yuichiro and Yoo, Chul-Moon",
    title = "{Primordial black holes and induced gravitational waves from a smooth crossover beyond standard model theories}",
    eprint = "2311.17760",
    archivePrefix = "arXiv",
    primaryClass = "astro-ph.CO",
    doi = "10.1103/PhysRevD.110.063521",
    journal = "Phys. Rev. D",
    volume = "110",
    number = "6",
    pages = "063521",
    year = "2024"
}

@article{Escriva:2024ivo,
    author = "Escriv{\`a}, Albert and Inui, Ryoto and Tada, Yuichiro and Yoo, Chul-Moon",
    title = "{LISA forecast on a smooth crossover beyond the standard model through the scalar-induced gravitational waves}",
    eprint = "2404.12591",
    archivePrefix = "arXiv",
    primaryClass = "astro-ph.CO",
    doi = "10.1103/PhysRevD.111.023528",
    journal = "Phys. Rev. D",
    volume = "111",
    number = "2",
    pages = "023528",
    year = "2025"
}

@article{Pritchard:2025pcn,
    author = "Pritchard, Xavier and Starbuck, Matthew and Leung, Wingfung",
    title = "{Beyond Standard Model equation of state and primordial black holes}",
    eprint = "2510.19629",
    archivePrefix = "arXiv",
    primaryClass = "astro-ph.CO",
    doi = "10.1088/1475-7516/2026/02/071",
    journal = "JCAP",
    volume = "02",
    pages = "071",
    year = "2026"
}

@article{Punturo:2010zz,
    author = "Punturo, M. and others",
    editor = "Ricci, Fulvio",
    title = "{The Einstein Telescope: A third-generation gravitational wave observatory}",
    doi = "10.1088/0264-9381/27/19/194002",
    journal = "Class. Quant. Grav.",
    volume = "27",
    pages = "194002",
    year = "2010"
}

@unpublished{Baker:2019nia,
    author = "Baker, John and others",
    title = "{The Laser Interferometer Space Antenna: Unveiling the Millihertz Gravitational Wave Sky}",
    eprint = "1907.06482",
    archivePrefix = "arXiv",
    primaryClass = "astro-ph.IM",
    reportNumber = "FERMILAB-PUB-19-436-A",
    month = "7",
    year = "2019"
}

@article{Coleman:1967ad,
    author = "Coleman, Sidney R. and Mandula, J.",
    editor = "Zichichi, A.",
    title = "{All Possible Symmetries of the S Matrix}",
    doi = "10.1103/PhysRev.159.1251",
    journal = "Phys. Rev.",
    volume = "159",
    pages = "1251--1256",
    year = "1967"
}

@article{Fayet:1977yc,
    author = "Fayet, Pierre",
    title = "{Spontaneously Broken Supersymmetric Theories of Weak, Electromagnetic and Strong Interactions}",
    reportNumber = "LPTENS 77/11",
    doi = "10.1016/0370-2693(77)90852-8",
    journal = "Phys. Lett. B",
    volume = "69",
    pages = "489",
    year = "1977"
}

@article{Farrar:1978xj,
    author = "Farrar, Glennys R. and Fayet, Pierre",
    title = "{Phenomenology of the Production, Decay, and Detection of New Hadronic States Associated with Supersymmetry}",
    reportNumber = "CALT-68-648",
    doi = "10.1016/0370-2693(78)90858-4",
    journal = "Phys. Lett. B",
    volume = "76",
    pages = "575--579",
    year = "1978"
}

@article{Kaplan:1983fs,
    author = "Kaplan, David B. and Georgi, Howard",
    title = "{SU(2) x U(1) Breaking by Vacuum Misalignment}",
    reportNumber = "HUTP-83/A069",
    doi = "10.1016/0370-2693(84)91177-8",
    journal = "Phys. Lett. B",
    volume = "136",
    pages = "183--186",
    year = "1984"
}

@article{Kaplan:1983sm,
    author = "Kaplan, David B. and Georgi, Howard and Dimopoulos, Savas",
    title = "{Composite Higgs Scalars}",
    reportNumber = "HUTP-83/A079",
    doi = "10.1016/0370-2693(84)91178-X",
    journal = "Phys. Lett. B",
    volume = "136",
    pages = "187--190",
    year = "1984"
}

@article{Weinberg:2003ur,
    author = "Weinberg, Steven",
    title = "{Damping of tensor modes in cosmology}",
    eprint = "astro-ph/0306304",
    archivePrefix = "arXiv",
    reportNumber = "UTTG-02-03",
    doi = "10.1103/PhysRevD.69.023503",
    journal = "Phys. Rev. D",
    volume = "69",
    pages = "023503",
    year = "2004"
}

@article{Watanabe:2006qe,
    author = "Watanabe, Yuki and Komatsu, Eiichiro",
    title = "{Improved Calculation of the Primordial Gravitational Wave Spectrum in the Standard Model}",
    eprint = "astro-ph/0604176",
    archivePrefix = "arXiv",
    doi = "10.1103/PhysRevD.73.123515",
    journal = "Phys. Rev. D",
    volume = "73",
    pages = "123515",
    year = "2006"
}

@article{Husdal:2016haj,
    author = "Husdal, Lars",
    title = "{On Effective Degrees of Freedom in the Early Universe}",
    eprint = "1609.04979",
    archivePrefix = "arXiv",
    primaryClass = "astro-ph.CO",
    doi = "10.3390/galaxies4040078",
    journal = "Galaxies",
    volume = "4",
    number = "4",
    pages = "78",
    year = "2016"
}

@article{NANOGrav:2023gor,
    author = "Agazie, Gabriella and others",
    collaboration = "NANOGrav",
    title = "{The NANOGrav 15 yr Data Set: Evidence for a Gravitational-wave Background}",
    eprint = "2306.16213",
    archivePrefix = "arXiv",
    primaryClass = "astro-ph.HE",
    doi = "10.3847/2041-8213/acdac6",
    journal = "Astrophys. J. Lett.",
    volume = "951",
    number = "1",
    pages = "L8",
    year = "2023"
}

@article{Cole:2022xqc,
    author = "Cole, Philippa S. and Gow, Andrew D. and Byrnes, Christian T. and Patil, Subodh P.",
    title = "{Smooth vs instant inflationary transitions: steepest growth re-examined and primordial black holes}",
    eprint = "2204.07573",
    archivePrefix = "arXiv",
    primaryClass = "astro-ph.CO",
    doi = "10.1088/1475-7516/2024/05/022",
    journal = "JCAP",
    volume = "05",
    pages = "022",
    year = "2024"
}

@article{Escriva:2022bwe,
    author = "Escriv{\`a}, Albert and Bagui, Eleni and Clesse, Sebastien",
    title = "{Simulations of PBH formation at the QCD epoch and comparison with the GWTC-3 catalog}",
    eprint = "2209.06196",
    archivePrefix = "arXiv",
    primaryClass = "astro-ph.CO",
    doi = "10.1088/1475-7516/2023/05/004",
    journal = "JCAP",
    volume = "05",
    pages = "004",
    year = "2023"
}

@article{Franciolini:2022tfm,
    author = "Franciolini, Gabriele and Musco, Ilia and Pani, Paolo and Urbano, Alfredo",
    title = "{From inflation to black hole mergers and back again: Gravitational-wave data-driven constraints on inflationary scenarios with a first-principle model of primordial black holes across the QCD epoch}",
    eprint = "2209.05959",
    archivePrefix = "arXiv",
    primaryClass = "astro-ph.CO",
    doi = "10.1103/PhysRevD.106.123526",
    journal = "Phys. Rev. D",
    volume = "106",
    number = "12",
    pages = "123526",
    year = "2022"
}

@article{Aoki:2006we,
    author = "Aoki, Y. and Endrodi, G. and Fodor, Z. and Katz, S. D. and Szabo, K. K.",
    title = "{The Order of the quantum chromodynamics transition predicted by the standard model of particle physics}",
    eprint = "hep-lat/0611014",
    archivePrefix = "arXiv",
    doi = "10.1038/nature05120",
    journal = "Nature",
    volume = "443",
    pages = "675--678",
    year = "2006"
}

@article{HotQCD:2018pds,
    author = "Bazavov, A. and others",
    collaboration = "HotQCD",
    title = "{Chiral crossover in QCD at zero and non-zero chemical potentials}",
    eprint = "1812.08235",
    archivePrefix = "arXiv",
    primaryClass = "hep-lat",
    doi = "10.1016/j.physletb.2019.05.013",
    journal = "Phys. Lett. B",
    volume = "795",
    pages = "15--21",
    year = "2019"
}

@article{Borsanyi:2020fev,
    author = "Borsanyi, Szabolcs and Fodor, Zoltan and Guenther, Jana N. and Kara, Ruben and Katz, Sandor D. and Parotto, Paolo and Pasztor, Attila and Ratti, Claudia and Szabo, Kalman K.",
    title = "{QCD Crossover at Finite Chemical Potential from Lattice Simulations}",
    eprint = "2002.02821",
    archivePrefix = "arXiv",
    primaryClass = "hep-lat",
    doi = "10.1103/PhysRevLett.125.052001",
    journal = "Phys. Rev. Lett.",
    volume = "125",
    number = "5",
    pages = "052001",
    year = "2020"
}

@article{Borsanyi:2013bia,
    author = "Borsanyi, Szabocls and Fodor, Zoltan and Hoelbling, Christian and Katz, Sandor D. and Krieg, Stefan and Szabo, Kalman K.",
    title = "{Full result for the QCD equation of state with 2+1 flavors}",
    eprint = "1309.5258",
    archivePrefix = "arXiv",
    primaryClass = "hep-lat",
    doi = "10.1016/j.physletb.2014.01.007",
    journal = "Phys. Lett. B",
    volume = "730",
    pages = "99--104",
    year = "2014"
}

@article{HotQCD:2014kol,
    author = "Bazavov, A. and others",
    collaboration = "HotQCD",
    title = "{Equation of state in ( 2+1 )-flavor QCD}",
    eprint = "1407.6387",
    archivePrefix = "arXiv",
    primaryClass = "hep-lat",
    reportNumber = "BNL-105928-2014-JA",
    doi = "10.1103/PhysRevD.90.094503",
    journal = "Phys. Rev. D",
    volume = "90",
    pages = "094503",
    year = "2014"
}

@article{Abuali:2025tbd,
    author = "Abuali, Ahmed and Bors{\'a}nyi, Szabolcs and Fodor, Zolt{\'a}n and Jahan, Johannes and Kahangirwe, Micheal and Parotto, Paolo and P{\'a}sztor, Attila and Ratti, Claudia and Shah, Hitansh and Trabulsi, Seth A.",
    title = "{New 4D lattice QCD equation of state: Extended density coverage from a generalized T' expansion}",
    eprint = "2504.01881",
    archivePrefix = "arXiv",
    primaryClass = "hep-lat",
    doi = "10.1103/2dmh-26yh",
    journal = "Phys. Rev. D",
    volume = "112",
    number = "5",
    pages = "054502",
    year = "2025"
}

@article{Borsanyi:2025dyp,
    author = "Borsanyi, Szabolcs and Fodor, Zoltan and Guenther, Jana N. and Parotto, Paolo and Pasztor, Attila and Ratti, Claudia and Vovchenko, Volodymyr and Wong, Chik Him",
    title = "{Lattice QCD constraints on the critical point from an improved precision equation of state}",
    eprint = "2502.10267",
    archivePrefix = "arXiv",
    primaryClass = "hep-lat",
    doi = "10.1103/rj6r-dmg9",
    journal = "Phys. Rev. D",
    volume = "112",
    number = "11",
    pages = "L111505",
    year = "2025"
}

@article{Abe:2023yrw,
    author = "Abe, Katsuya T. and Tada, Yuichiro",
    title = "{Translating nano-Hertz gravitational wave background into primordial perturbations taking account of the cosmological QCD phase transition}",
    eprint = "2307.01653",
    archivePrefix = "arXiv",
    primaryClass = "astro-ph.CO",
    doi = "10.1103/PhysRevD.108.L101304",
    journal = "Phys. Rev. D",
    volume = "108",
    number = "10",
    pages = "L101304",
    year = "2023"
}

@article{Domenech:2021ztg,
    author = "Dom{\`e}nech, Guillem",
    title = "{Scalar Induced Gravitational Waves Review}",
    eprint = "2109.01398",
    archivePrefix = "arXiv",
    primaryClass = "gr-qc",
    doi = "10.3390/universe7110398",
    journal = "Universe",
    volume = "7",
    number = "11",
    pages = "398",
    year = "2021"
}

@unpublished{Zeng:2025cer,
    author = "Zeng, Xiang-Xi and Ning, Zhuan and Cai, Rong-Gen and Wang, Shao-Jiang",
    title = "{Scalar-induced gravitational waves with non-Gaussianity up to all orders}",
    eprint = "2508.10812",
    archivePrefix = "arXiv",
    primaryClass = "astro-ph.CO",
    month = "8",
    year = "2025"
}

@article{Li:2025met,
    author = "Li, Jun-Peng and Wang, Sai and Zhao, Zhi-Chao and Kohri, Kazunori",
    title = "{Isotropy, anisotropies and non-Gaussianity in the scalar-induced gravitational-wave background: diagrammatic approach for primordial non-Gaussianity up to arbitrary order}",
    eprint = "2505.16820",
    archivePrefix = "arXiv",
    primaryClass = "astro-ph.CO",
    reportNumber = "KEK-TH-2724, KEK-Cosmo-0381",
    doi = "10.1088/1475-7516/2026/05/064",
    journal = "JCAP",
    volume = "05",
    pages = "064",
    year = "2026"
}
\let\addcontentsline\oldaddcontentsline
\end{document}